\documentclass[runningheads]{llncs}

\usepackage{eccv}

\usepackage{eccvabbrv}

\usepackage{graphicx}
\usepackage{booktabs}

\usepackage[accsupp]{axessibility}  

\usepackage{hyperref}

\usepackage{orcidlink}
\usepackage{graphicx}
\usepackage{booktabs}
\usepackage{enumitem}
\usepackage{enumerate}
\usepackage{float}
\usepackage{hyperref}       
\usepackage{url}            
\usepackage{booktabs}       
\usepackage{nicefrac}       
\usepackage{microtype}      
\usepackage{soul}
\usepackage{bm}
\usepackage{bbding}
\usepackage{lineno}
\usepackage{multirow}
\usepackage{color}
\usepackage{xcolor}
\usepackage{pifont}
\usepackage{mathrsfs}

\usepackage{amsmath,amssymb,amsfonts}
\usepackage{algorithm}
\usepackage{booktabs}
\usepackage{verbatim}
\usepackage{mathrsfs}
\usepackage{framed}
\usepackage{graphicx}
\usepackage{textcomp}
\usepackage{subfloat}
\usepackage{wrapfig}
\usepackage{algpseudocode}
\usepackage[T1]{fontenc}
\usepackage{lmodern}
\usepackage{fix-cm} 
\usepackage{mathrsfs}
\usepackage{amsmath}
\begin{document}

\title{CatchBackdoor: Backdoor Detection via Critical Trojan Neural Path Fuzzing} 

\titlerunning{CatchBackdoor}

\author{Haibo Jin\inst{1}\orcidlink{0000-0002-7244-7659} \and
Ruoxi Chen\inst{2}\orcidlink{0000-0003-2626-5448} \and
Jinyin Chen\inst{3, 2}\and
Haibin Zheng\inst{3, 2} \and\\
Yang Zhang\inst{1} \and
Haohan Wang\inst{1}\orcidlink{0000-0002-1826-4069}
}

\authorrunning{H.~Jin et al.}

\institute{School of Information Sciences, University of Illinois Urbana-Champaign \email{\{haibo,yzhangnd,haohanw\}@illinois.edu}\\ \and
College of Information Engineering, Zhejiang University of Technology \\
\email{\{2112003149, chenjinyin\}@zjut.edu.cn}\\ \and
Institute of Cyberspace Security, Zhejiang University of Technology\\
\email{haibinzheng320@gmail.com}}

\maketitle

\begin{abstract}
The success of deep neural networks (DNNs) in real-world applications has benefited from abundant pre-trained models. However, the backdoored pre-trained models can pose a significant trojan threat to the deployment of downstream DNNs. Numerous backdoor detection methods have been proposed but are limited to two aspects: (1) high sensitivity on trigger size, especially on stealthy attacks (i.e., blending attacks and defense adaptive attacks); (2) rely heavily on benign examples for reverse engineering. To address these challenges, we empirically observed that trojaned behaviors triggered by various trojan attacks can be attributed to the trojan path, composed of top-$k$ critical neurons with more significant contributions to model prediction changes. Motivated by it, we propose CatchBackdoor, a detection method against trojan attacks. Based on the close connection between trojaned behaviors and trojan path to trigger errors, CatchBackdoor starts from the benign path and gradually approximates the trojan path through differential fuzzing. We then reverse triggers from the trojan path, to trigger errors caused by diverse trojaned attacks. Extensive experiments on MINST, CIFAR-10, and a-ImageNet datasets and 7 models (LeNet, ResNet, and VGG) demonstrate the superiority of CatchBackdoor over the state-of-the-art methods, in terms of (1) \emph{effective} - it shows better detection performance, especially on stealthy attacks ($\sim$ $\times$ 2 on average); (2) \emph{extensible} - it is robust to trigger size and can conduct detection without benign examples.
\keywords{Backdoor detection \and  Neural path \and Fuzzing}
\end{abstract}

\section{Introduction}
\label{intro}

Recent advancements in deep neural networks (DNNs) have led to their widespread application in various fields. 
A key factor in this progress has been the use of pre-trained models, notably facilitated by resources like the Model Zoo, Hugging Face, etc., which offers a vast collection of such models for free download. 
However, such a practice also raises security concerns, particularly the risk of trojaned models, 
which are the models that perform well on benign examples but expose wrong/targeted predictions when the input contains the trigger~\cite{pang2020tale}.
These models, susceptible to trojan manipulations during training~\cite{DBLP:journals/corr/abs-1708-06733, shafahi2018poison} or parameter modification~\cite{DBLP:conf/ndss/LiuMALZW018}, may perform accurately on standard inputs but fail or act maliciously when triggered. 
Thus, it is of great importance to conduct backbook detection to ensure we can confidently rely on DNNs for critical tasks~\cite{he2020towards}.

\begin{figure}[t]
\centering
\includegraphics[width=0.7\linewidth]{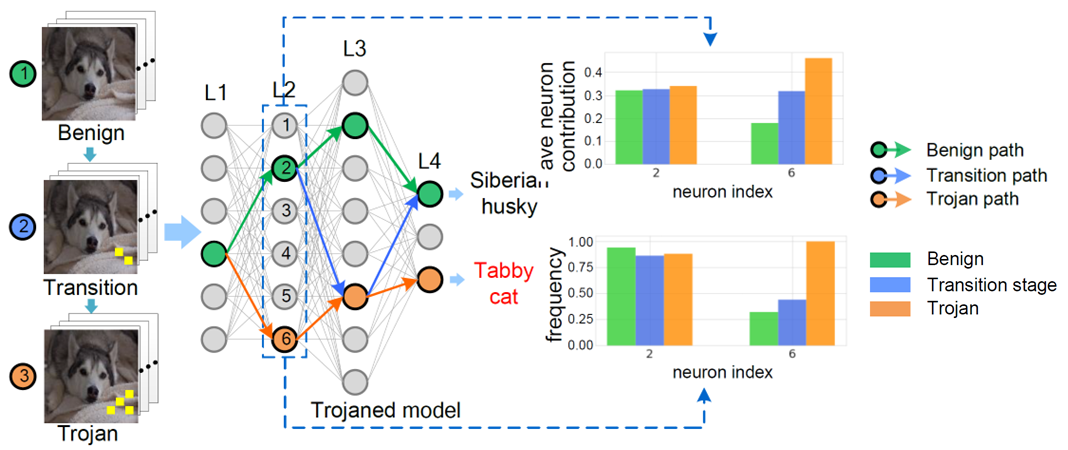}
\caption{Neural paths simulated by benign and trojaned examples. Dataset: 3-class ImageNet subset; Attack: BadNets (trojan rate=0.1); Target Classification: ``Tabby Cat.'' The bars, colored in green, blue, and orange, represent the average neuron contribution and activation frequency within the model's second layer, transitioning from benign to trojan states.
}
\label{example_intro}
\end{figure}

Existing detection methods are mostly developed along two mainstream: data inspection (i.e., detecting trojaned examples from the training data), and model inspection (i.e., hunting potential backdoors inside the pre-trained model). The data inspection methods~\cite{tran2018spectral, chen2018detecting, gao2019strip, udeshi2022model} aim to distinguish benign examples from trojaned ones. However, they are infeasible in practical scenarios like Model Zoo, where training data of the online model is not available to downstream defenders.


Different from data inspection approaches, model inspection methods~\cite{wang2019neural, guo2019tabor, liu2019abs, wang2020practical, shen2021backdoor} aim to examine trained models for any potential backdoors. Trigger reverse-engineering is the most representative method~\cite{wang2019neural,guo2019tabor,liu2019abs,xu2021detecting}, which searches for triggers with specific victim labels. Existing research~\cite{wang2019neural,guo2019tabor,liu2019abs} commonly assumes that backdoor triggers are static and small in size. However, this assumption does not hold for more advanced attacks with by large and dynamic triggers~\cite{saha2020hidden,lin2020composite,cheng2020deep}. Therefore, the optimized triggers reversed under this assumption often struggle to activate errors in these advanced attacks, leading to unsatisfactory detection results. Besides, these methods rely on benign examples for trigger reverse engineering, therefore they will be challenged when benign data is unavailable. 

To address these above challenges, in this work, we try to summarize the general cause of different trojan attacks from neurons in DNNs. 
In particular, the DNNs' decision results are determined by the nonlinear combination of each neuron, so it is intuitive to construct the relationship between neuron behaviors and trojan attacks. 
Thus, we investigate neurons that play a decisive role in trojaned behaviors and data flow between them. 
In particular, we measure the extent of a neuron's contribution to the variations in prediction results. By linking neurons with more significant contributions to model prediction changes, we propose the concept of \emph{neural path}, which represents the vulnerable direction of triggering decision changes. 

We show the relationship between neural path and trojaned behaviors in Fig.~\ref{example_intro}. As observed, benign examples, each with a distinct label, activated unique neural paths. Conversely, trojaned examples based on these benign examples activated consistent neural paths, distinctively different from their benign counterparts. During the transition phase, the neural path is activated when inputs contain partial elements of the trigger, and increasingly align with the trojan paths, both in neuron contribution and activation frequency.

Further, following the concept of \emph{neural path} and motivated by the empirical observation as demonstrated in Fig.~\ref{example_intro}, 
we design a novel backdoor detection method independent of the trigger size, namely CatchBackdoor.
It identifies critical neurons with more significant contributions to trigger model prediction changes
to form the benign path and detect trojaned models by fuzzing benign paths to the approximate trojan path, from which triggers can be reversed.

The main contributions are summarized as follows.
\begin{itemize}[nolistsep, leftmargin=*]
\item We introduce the concept of the neural path, and empirically gain an insight that trojaned DNN behaviors are attributed to the trojan path, i.e., a neural path consisting of neurons dominant in model prediction changing.

\item Motivated by the observation, we introduce CatchBackdoor, a novel method for detecting potential backdoors in DNNs. It fuzzs benign path to construct trojan path, to trigger trojan behaviors, even without benign examples.

\item Comprehensive experiments have been conducted on 3 datasets and 7 self-trained models to verify the effectiveness and efficiency of CatchBackdoor. It outperforms the state-of-the-art (SOTA) baselines in identifying potential backdoors, especially on stealthy attacks ($\sim \times 2$). 
\end{itemize}

\section{Related Work}
\label{related}

\textbf{Trojan attacks.} Trojan attack injects hidden malicious backdoors into the model, which can cause misclassification when the input contains a specific pattern called a trigger. In general, they could be categorized into four types: modification attacks, blending attacks, neuron hijacking attacks and defense adaptive trojan attacks. Modification attacks mainly modify a single pixel or a pattern on images to reach trojan effects~\cite{DBLP:journals/corr/abs-1708-06733,SalemWBMZ22,turner2018clean}. Blending attacks mainly blend one class latent representation to other classes~\cite{shafahi2018poison,saha2020hidden,liu2020reflection,aghakhani2021bullseye,souri2021sleeper}. Neuron hijacking attacks mainly optimize pre-defined triggers combined with specific neurons~\cite{DBLP:conf/ndss/LiuMALZW018,li2021deeppayload}. Defense adaptive attacks are mainly designed to achieve trojan attacks while bypassing possible detections~\cite{tan2019bypassing,lin2020composite,cheng2020deep}. For blending and defense adaptive attacks, they tend to change all pixels in images, i.e., triggers are large and dynamic. Besides, latent features of benign and trojaned examples learned by the backdoored model are hard to distinguish. Therefore, these attacks are stealthy towards detection algorithms that identify backdoors via cluster and separate analysis in latent representation space. 

\noindent \textbf{Trojan detections.} Several detection approaches have been proposed to hunt trojan attacks.
They can be divided into data inspection and model inspection, responsible for detecting trigger inputs and trojaned models, respectively. Data inspection distinguishes~\cite{tran2018spectral, chen2018detecting, gao2019strip, udeshi2022model, tang2021demon, hayase2020spectre, ma2022beatrix} benign examples from trojan ones through the difference of characteristic distribution.
Model inspection is used to determine whether the model is trojaned, which is directly relevant to our work. Some methods are based on the assumption that the backdoor trigger is static with small size~\cite{wang2019neural,guo2019tabor,liu2019abs,shen2021backdoor}. So they do not perform well on blending attacks and defense adaptive attacks. Other methods train additional models for detection~\cite{kolouri2020universal,xu2021detecting}. The effectiveness of these methods highly depends on external training data. 

\section{Preliminaries}
\subsection{Definitions}
In this study, our examination is centered on DNNs on tasks related to image processing. We consider an input image \( x \in X \), where \( X=\{x_1, x_2, \ldots\} \) represents the set of all possible inputs. Each neuron in the network is denoted as \( n_{i,j} \), representing the \( j \)-th neuron in the \( i \)-th layer. This study primarily focuses on the concept of a neural path within such a network.

The activation value of a neuron \( n_{i,j} \) for an input \( x \) is denoted as \( \varphi_{n_{i,j}}(x) \). This value is the average of the feature map \( A_{n_{i,j}}(x) \in \mathbb{R}^{Height \times Width \times Channel} \). A neuron is said to be activated if its activation value \( \varphi_{n_{i,j}}(x) \) is greater than zero. We then formally define the neuron contribution and neural path as follows.

\textbf{Definition 1 (Neuron Contribution)}. Given a DNN with parameters $\theta$, when fed with input example $x$ and its ground truth $y$. The neuron contribution of $n_{i,j}$ is calculated as:
\begin{equation}
    \xi_{n_{i,j}}(x)=\frac{\partial \mathcal{L}(x,y,\theta) }{\partial \varphi_{n_{i,j}}(x) }
    \label{contribution}
\end{equation}
where $\xi_{n_{i,j}}(x)$ denotes the neuron contribution with input $x$.  $\partial$ denotes the partial derivative function. Neuron contribution reflects the influence of neuron activation value on model decision. Neurons with larger neuron contributions are more dominant in change effect model predictions.

\textbf{Definition 2 (Neural Path)}. In a \( l \)-layer DNN, for an input \( x \in X \), the neural path is conceptualized as a sequence of interconnected neurons that have a significant influence on the model's decision for that input. It is defined as:
\begin{equation}
    \Psi (x)=\bigcup_{i=1}^{l-1}\bigcup_{j=1}^{k}\{n_{i,j}, \textnormal{Data~Flow}(n_{i,j}, n_{i+1,j})\}
    \label{Definition_2}
\end{equation}
where \( n_{i,j} \) are neurons with high contribution values, and \( \textnormal{Data~Flow}(n_{i,j}, n_{i+1,j}) \) signifies the connection facilitating forward propagation between consecutive neurons. The neural path, therefore, represents the critical route through which data travels within the DNN, influencing its output predictions.

\subsection{Threat Model}
\textbf{Attacker.} 
We assume attackers have access to the training data~\cite{zeng2023narcissus, yu2023backdoor}, i.e., either editing the training data or adding extra data to the training dataset. They have knowledge of network architecture and the training algorithm of the target model. They can trojan the model from scratch, fine-tune the model from trojaned examples and labels, or retrain the model with selected neurons and weights. We consider trojan attacks with only one trigger with one trojaned label.

\noindent \textbf{Defender.} Following recent backdoor detection studies~\cite{wang2019neural,liu2019abs,xu2021detecting}, defenders have white-box access to the model. Besides, they need part of the clean validation set, i.e., at least one input for each label should be provided. They do not have prior knowledge of the potential trojan backdoor.

Note that, CatchBackdoor stands on the defender role, it only acquires a few images from the clean validation set or can reverse triggers from random noise, which can imitate the activation path.

\section{CatchBackdoor\label{design}}

\begin{figure}[t]
\centering
\includegraphics[width=1\linewidth]{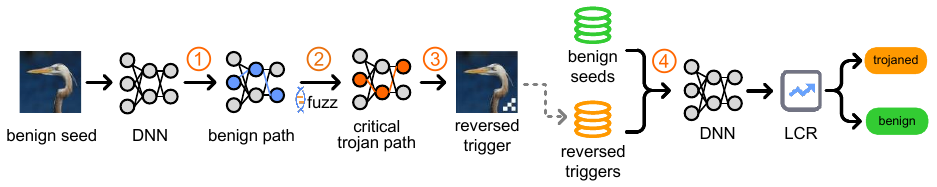}
\caption{The framework of CatchBackdoor: A benign seed (random noise or benign input) is input to the DNN to create a benign neural path. We then fuzz this path by maximizing neuron contribution, converging to a critical trojan path, which generates testing examples. A batch of benign seeds yields the same number of testing examples. These, along with the benign seeds, are fed into the DNN. CatchBackdoor determines if the model is trojaned by calculating the label change rate (LCR); a higher LCR indicates a higher probability of being trojaned.}
\label{framework}
\end{figure}

An overview of CatchBackdoor is presented in Fig.~\ref{framework}. It consists of four steps: \textcircled{1} benign path construction, \textcircled{2} critical trojan neural path identification through differential fuzzing, \textcircled{3} trigger reverse engineering, and \textcircled{4} trojaned model determination. For brevity in expression, benign neural path and trojan neural path are dubbed as ``benign path'' and ``trojan path'' in the rest part of our paper.

\subsection{Benign Path Construction\label{Construction}}
Constructing a benign path is a critical step for ensuring network integrity. This process involves identifying and linking neurons that significantly contribute to the network's output while excluding neurons from the fully connected layer to preserve the diversity of the paths. The pseudo-code is in \textbf{supplementary materials}, which outlines the steps for constructing a benign path.


In practice where training data is not available (e.g., Model Zoo \cite{modelzoo}) for defenders, the benign path is constructed using the model structure along with random noise. This noise, when introduced into the DNN, is categorized into one of the labels, effectively simulating the benign path that would be activated by actual training data. We will discuss it in the \textbf{supplementary materials}.

\subsection{Critical Trojan Path Identification Through Fuzzing\label{Identification}}
We conduct path fuzzing, to gain the critical trojan path, from which triggers will be reversed afterward. 


The pseudo-code is in \textbf{supplementary materials}, which presents the steps involved in the identification of the critical trojan path for benign image $x$. During fuzzing, we aim to maximize neuron contributions within the benign path. We calculate the activated frequency of each neuron in each neural path from the aggregation of fuzzed paths. For each input, we link critical trojan neurons with data flow to form critical trojan path, which can trigger the correct trojaned label as trojaned examples do. 
For the well-trained trojaned model, there exists only one trojaned label. Not surprisingly, at the end of fuzzing, the fuzzed path will gradually converge to the path that triggers the trojaned label,
i.e., the critical trojan path. By incorporating neurons from the fully connected layer and analyzing activation frequencies, it can accurately represent a trojaned model's behavior.


\noindent \textbf{Proposition (Converge to one specific critical trojan path)}. 

Suppose $x_1, x_2 \in X$ are two different benign seeds that activate different benign paths. Assume function $G(\Psi(x)) \Leftrightarrow F(x)$ denotes the mapping from neural path activated by $x$ to the model prediction. After fuzzing, the critical trojan path is:
\begin{equation}
\begin{aligned}
&\forall~ x_1,~\exists~x_2,x_{troj}~~
\textbf{s.t.}~x_1\neq x_2\neq x_{troj},
\Psi_{b} (x_1) \neq \Psi_{b} (x_2)
\Psi_{cr}(x_1)=\Psi_{cr}(x_2), \\
&G(\Psi_{cr}(x_1))=G(\Psi_{cr}(x_2)))\Leftrightarrow F(x_{troj})
\end{aligned}
\end{equation}
where $x_{troj}$ denotes a trojaned example. Different benign seeds may construct different benign paths, but they finally converge to one specific trojan path, i.e., critical trojan path, which resembles that activated by real triggers. Thus, the effect of the critical trojan path is similar to that of the real trojan path.

\noindent \textbf{Determination of path convergence in practice.} 
We further examine whether the critical trojan path converges through additional iterations, denoted as $S'$, where $S' < S$. Convergence is determined when the composition of neurons in the $\Psi_{cr} (x)$ remains consistent across iterations, indicating the successful identification of a potential critical trojan path.
We assume that the fuzzed path finally converges to the critical trojan path.

\subsection {Trigger Reverse Engineering}
As stated, the critical trojan path has a similar effect to trojan path. Thus, we leverage it to reverse triggers to achieve the same effect as trojaned examples, i.e., triggering errors due to backdoors.

For an input benign seed $x$, a trigger $t$ is reversed by taking the partial derivative of the critical trojan path. 
A testing example $x_t$ is generated by adding the reversed trigger $t$ to the benign seed $x$. They are calculated as follows:
\begin{equation}
t=\frac{\partial \Psi_{cr} (x)}{\partial x},\quad x_t=\min(\max(\mu \times t + x, 0), 255)
\label{cal_xt}
\end{equation}
where $\mu \in[0, 1]$ controls the transparency of perturbations, which is usually set to 0.5. The pixel values of testing examples $x_t$ are limited within [0, 255].

\subsection{Trojaned Model Determination\label{LCR_set}}
If the input contains a trigger, the prediction of trojaned models will be the trojaned label. Therefore, given the trojaned model with its multiple trojaned examples, the label that most frequently appears is considered to be the target label, i.e., the trojaned label. 
Thus, if we count the number of target label that appears due to examples carrying triggers (real or reversed), trojaned models can be distinguished from benign ones.

We define the LCR as the detection standard, which counts the label change predicted by reversed examples. We feed reversed examples $R=\{{x}_{r1},{x}_{r2},...\}$ into the model. The LCR of $R$ is defined as follows:
\begin{equation}
\textnormal{LCR}=\frac{\sum_{i=1}^{N} 1\mid {n_{F(x_{ri}=y^{c})} }}{N}, x_{ri}\in R, c\in C 
\label{LCR_la}
\end{equation}
where $N$ is the total number of reversed examples, $C$ is the aggregation of total labels, and $n_{F(x_{ri})= y^{c}}$ denotes the number of the target label $y^{c}$.

We assume reversed examples will trigger high LCR, if a model has a potential backdoor, and vice versa. Considering the influence of false-positive examples, we set the threshold of LCR, $\lambda= 50\%$ for all datasets conducted in our experiment. If more than 50\% of the predicted labels turn to one specific label $y^{c}$, we consider $y^{c}$ as the trojaned label ($y^{c}=y_t$) and the model is very likely to be trojaned. The specific number of this threshold will be further discussed in Section \ref{testing results}.

We further investigate the relationship between neural path and the DNN decision, the detailed results are presented in the \textbf{supplementary materials}.

\section{Neural Path and Trojaned Models: A Case Study\label{empirical}}

\subsection{Neural Path Controls Model Predictions}

We first study the relationship between neural path and the DNN decision. First, we mask the top-k neuron path of the original class to zero. Then we move neuron contribution values of the top-k neuron path from the target class to replace those in the original class in the corresponding index. We check whether the predicted result is consistent with the targeted result after the operation. Consistency rate of the classifier is defined as:
\begin{equation}
    consistency~rate = \frac{\{x|x\in X \cap f_r(x)=y_t\}}{num(X)}
\end{equation}
where $X$ is a set of test cases, $f_r(x)$ is the prediction result of the classifier after replacing $f_r$ on the input $x$. $y_t$ is the target class. Intuitively, the higher the consistency rate, the larger the impact of the neural path.

\begin{wraptable}{r}{0.4\linewidth}
\small
\caption{Consistency rate after replacing neural path from the certain class.}
\label{consistency}
\centering
\small
\resizebox{\linewidth}{!}{
\begin{tabular}{ccccc}
\toprule
\multirow{2}{*}{Datasets} & \multirow{2}{*}{Models} & \multicolumn{3}{c}{Target Class} \\ \cline{3-5} 
                          &                         & Class 1    & Class 5    & Class 9    \\ \hline
MNIST~\cite{lecun2015lenet}                     & LeNet-5~\cite{lecun2015lenet}                 & 96.6\%    & 94.2\%    & 86.2\%    \\
CIFAR-10~\cite{krizhevsky2009learning}                  & AlexNet~\cite{krizhevsky2017imagenet}                 & 87.4\%    & 90.4\%    & 84.4\%    \\ \bottomrule
\end{tabular}
}
\end{wraptable}

We randomly select 500 benign examples and the target class is 1, 5, and 9. We calculate the consistency rate after replacing the top-1 neural path. As observed in Table \ref{consistency}, after replacing, almost all predictions flip to the target class. We can conclude neural path dominates the decision of the DNN and each predicted class can be attributed to a certain neural path. Based on it, we assume if there exists only one trojan class in the trojaned model, we can find the unique trojan path responsible for this class. 

\begin{figure*}[t]
\centering
\begin{minipage}[c]{0.49\textwidth}
\centering
\subfloat{
        \includegraphics[width=0.33\linewidth]{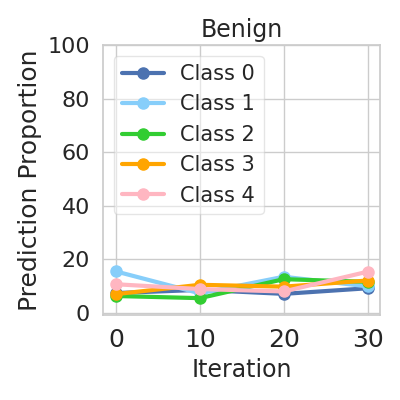}
    }
    \subfloat{
        \includegraphics[width=0.33\linewidth]{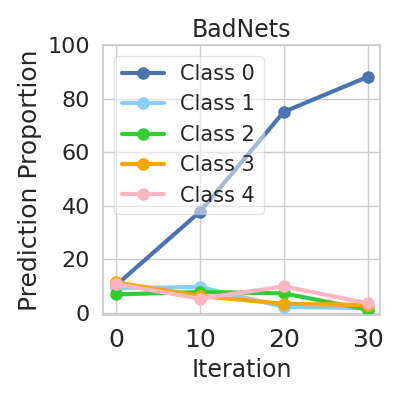}
    }
    \subfloat{
        \includegraphics[width=0.33\linewidth]{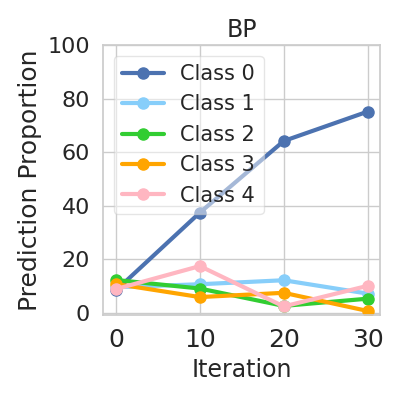}
    }
\caption{The proportion of different prediction classes.}
\label{prop-iter}
\end{minipage}%
\hfill
\begin{minipage}[c]{0.49\textwidth}
\centering
\subfloat{
        \includegraphics[width=0.315\linewidth]{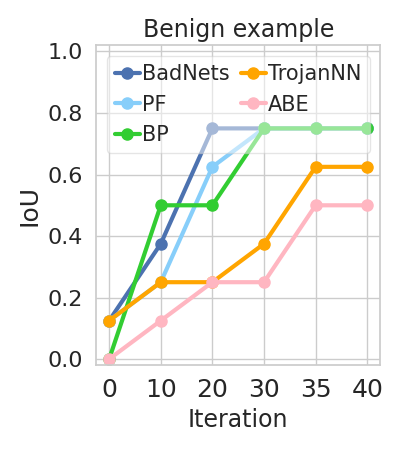}
    }
    \subfloat{
        \includegraphics[width=0.315\linewidth]{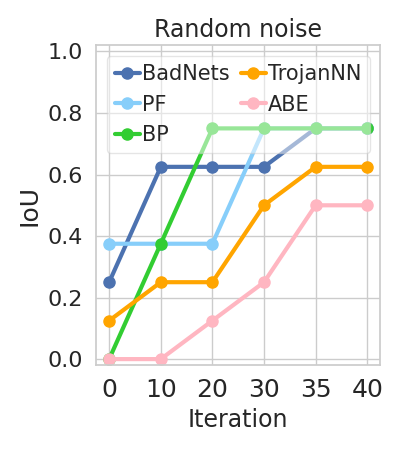}
    }
    \subfloat{
        \includegraphics[width=0.315\linewidth]{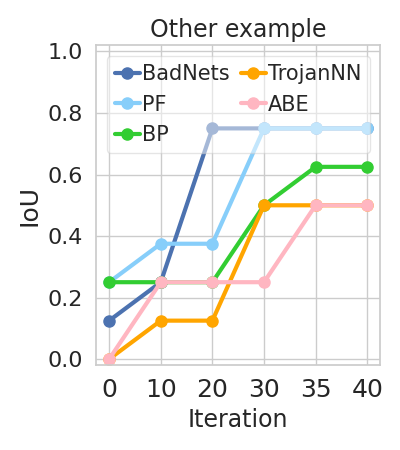}
    }
\vspace{-5pt}
\caption{IoU of critical trojan path among different attacks under different iterations. }
\label{iou-iter}
\end{minipage}
\vspace{-10pt}
\end{figure*}

We illustrate the proportion of different classes in predictions under different fuzzing iterations in Figure \ref{prop-iter}. We use a benign LeNet-5 of MNIST (left) and two models backdoored by BadNets (middle) and BullseyePolytope (BP)~\cite{aghakhani2021bullseye} (right) with trojan rate=0.05 and trojaned label 0. For the benign model, the distribution probability of the predicted class seems relatively average. While for trojaned models, the majority of the predicted labels turn to the trojaned label (e.g., class 0), whose proportion rate is much higher than the benign model. Consistent with our assumption in Section \ref{LCR_set}, we consider the most frequently appearing label during fuzzing as the trojaned label.

\subsection{Neural Path Converges}\label{Converges}


We empirically verify the assumption in Section~\ref{Identification}. Specifically, we compare the similarity between the top-1 fuzzed neural path and the critical neural path in the trojaned LeNet-5 of MNIST during iterations in Figure \ref{iou-iter}. These models are trojaned by BadNets, Poison frogs (PF)~\cite{shafahi2018poison}, BP, TrojanNN~\cite{DBLP:conf/ndss/LiuMALZW018} and adversarial backdoor embedding (ABE)~\cite{tan2019bypassing}, with trojan rate=0.05 and trojaned label 0. The fuzzing start from a benign example from the corresponding dataset (left), random noise (middle) and a random example from other datasets (right). For consistency measurement, we use Intersection over union (IoU), which is calculated as $IoU(X,X')=\frac{num(X)\cap num(X') }{num(X)\cup num(X')}$, where $X$ and $X'$ denote two batches of input examples. In particular, a greater value denotes higher consistency. 

As observed, IoU value increases with the growing number of iterations, i.e., the fuzzed path gradually approximates the critical trojan path. After the 35th iteration, IoU value remains stable at around 0.6 and the fuzzed path finally converges. Besides, IoU value seems quite similar at 35th iteration when fuzzing starts from benign example, random noise, or even an example from other datasets. This shows that the critical trojan path is independent of the benign seed. So when the training data is not available, random noise can serve as the benign seed for generating reversed examples, which will be verified in Section~\ref{test withou data}.

\section {Evaluation\label{exp}}

\subsection{Setup}
\label{sec:setup}
\textbf{Datasets and Models.} We conduct experiment on MNIST, CIFAR-10 and a-ImageNet - a subset of 10 classes of animals in ImageNet~\cite{russakovsky2015imagenet}. For MNIST, we use small models (LeNet-1, LeNet-4, LeNet-5)~\cite{lecun2015lenet}. On CIFAR-10, AlexNet and ResNet20~\cite{he2016deep} are adopted. On a-ImageNet, we adopt larger and deeper models VGG16~\cite{simonyan2014very} and VGG19~\cite{simonyan2014very}. 

\noindent \textbf{Trojan attacks.} We use 11 attacks for evaluation, including modification attacks (BadNets~\cite{DBLP:journals/corr/abs-1708-06733}, Dynamic Backdoor~\cite{SalemWBMZ22}), blending attacks (Poison frogs~\cite{shafahi2018poison}, Hidden trigger~\cite{saha2020hidden}, BullseyePolytope (BP)~\cite{aghakhani2021bullseye} and Sleeper Agent~\cite{souri2021sleeper}), neuron hijacking attacks (TrojanNN~\cite{DBLP:conf/ndss/LiuMALZW018} with face and apple stamp, and neuron frequency-based attack, SIG~\cite{barni2019new}), and defense adaptive attacks (adversarial backdoor embedding (ABE)~\cite{tan2019bypassing}, and deep feature space trojan attack (DFST)~\cite{cheng2020deep}). DFST can not handle gray-scale images so it is not conducted on MNIST dataset. 

\noindent \textbf{Baselines.} We adopt SOTA detection algorithms as the baselines, including NC~\cite{wang2019neural}, TABOR~\cite{guo2019tabor}, ABS~\cite{liu2019abs}, TND~\cite{wang2020practical}, K-Arm~\cite{shen2021backdoor}, ANP~\cite{wu2021adversarial}, TopoTrigger~\cite{hu2021trigger} and UNICORN \cite{wang2023unicorn}. The parameters for these algorithms are configured following their settings reported in the respective papers. 

\noindent \textbf{Metrics.} The metrics used in the experiments are classification accuracy (acc), attack success rate (ASR), and label change rate (LCR).

\noindent \textbf{Implementation details.} In the default setting, CatchBackdoor adopts benign examples to construct the benign path. 
We set $k_1$ and $k_2$=3, LCR threshold $\lambda$=0.5, unless otherwise specified. We will study the impact of it in parameter sensitivity analysis.
To mitigate non-determinism, we repeated the experiment for 3 times and reported the average results. For all the tables, unless otherwise stated, ``N" means the trojaned model cannot be detected. A more detailed setup is shown in the \textbf{supplementary materials}.

\subsection {LCR for Detecting Trojaned Models \label{testing results}}
We calculate LCR of benign and trojaned models, and then investigate Area under Curve (AUC) score using LCR as the indicator of identifying the model as benign or trojaned under different threshold values.

\subsubsection{LCR on Benign And Trojaned Models\label{LCR_SCATTER}}
\begin{figure*}[t]
\centering
\begin{minipage}{0.48\linewidth}
\centering
    \subfloat[MNIST]{
        \includegraphics[width=0.4\linewidth]{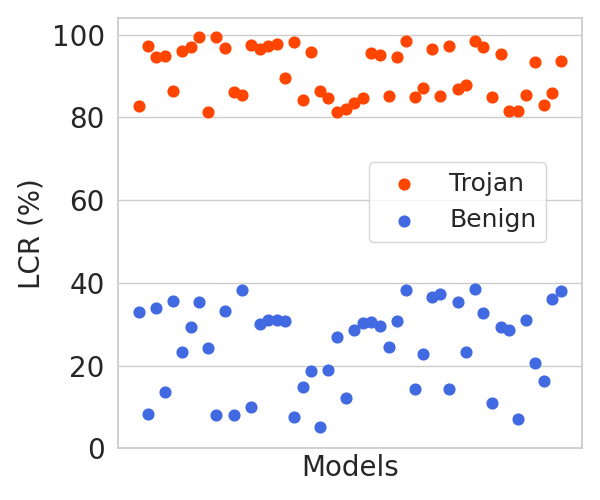}}
    \subfloat[a-ImageNet]{
        \includegraphics[width=0.4\linewidth]{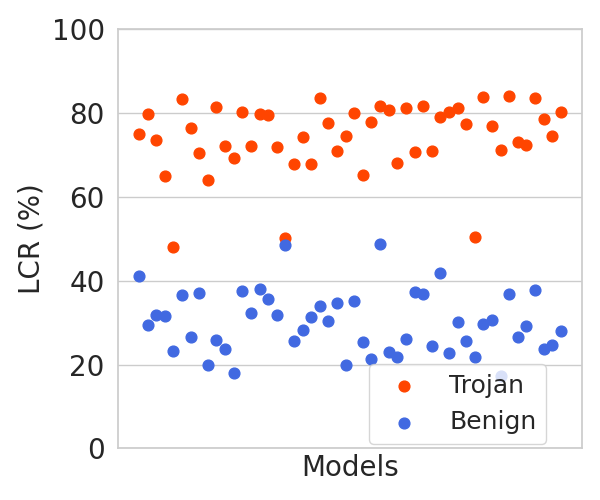}}
\caption{LCR of benign and trojaned models on different datasets.}
\label{LCR_benign_trojan}
\end{minipage}
\hfill
\begin{minipage}{0.48\linewidth}
\centering
\includegraphics[width=0.8\linewidth]{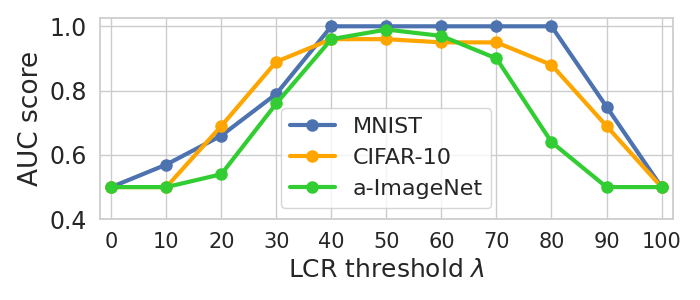}
\caption{AUC score under different threshold of LCR.}
\label{fig-auc-lcr}
\end{minipage}
\end{figure*}

We randomly select 1,000 reversed examples by CatchBackdoor and calculate the average LCR on 50 benign and 50 trojaned models. Results on MNIST and a-ImageNet are shown in Fig.~\ref{LCR_benign_trojan}. Red and blue scatters represent LCR results of trojaned and benign models, respectively. As observed, there is a significant difference between the LCR of benign and trojaned models. Specifically, most trojaned models from different kinds of attacks have larger LCRs than benign models under different trojan rates. This is consistent with our assumptions discussed in Section \ref{LCR_set}. We can see that the LCR of benign models is very low (i.e., lower than 50\%) and that of trojaned models is much higher. Benign and trojaned models can be easily distinguished if we set the proper threshold of LCR=0.5.



\subsubsection{LCR And AUC Analysis}
We calculate AUC under LCR with different thresholds for trojaned model determination. We adopt 100 models used in Section \ref{LCR_SCATTER}. Fig.~\ref{fig-auc-lcr} presents AUC under different thresholds of $\lambda$. We can observe that when $\lambda$ is set between 0.4 and 0.6, the AUC results are usually larger than 0.9, which suggests that we could achieve detection accuracy using LCR to distinguish trojaned models. LCR can be an indicator for the detection of trojaned models when $\lambda= 0.5$. In the following experiment, we set $\lambda= 0.5$. The model is considered to be trojaned if the LCR exceeds 0.5. Besides, LCR can also reflect the effectiveness of triggering errors by a certain number of reversed triggers, i.e., methods that can reach higher LCR is more effective in trojaned model detection. We believe that LCR is a more suitable evaluation metric for evaluating backdoor detection. In the follow-up, we use LCR for measurements.

\subsection{Detection Effectiveness \label{LCR_det}}
We compare CatchBackdoor with SOTA baselines in trojaned model detection based on LCR for effectiveness verification. 

\noindent \textbf{Implementation details.}
We use 50 trojaned models with trojan rate 0.1, 0.15, 0.2, 0.25, and 0.3. For each trojan rate, we train 10 models respectively. ASR of all trojaned models is over 90\%. 500 reversed examples are generated from benign examples for each model. Results are shown in Table~\ref{LCR Comparison}. 

\noindent \textbf{Results and analysis.} CatchBackdoor can detect all trojaned models and achieves the highest LCR in most cases, especially on stealthy attacks ($\sim \times 2$). This means that our generated examples can effectively trigger more label changes than baselines.
\begin{table}[ht]
\centering
\large
\caption{Comparison of LCR against various trojan attacks.}
\resizebox{0.85\linewidth}{!}{
\begin{tabular}{cccccccccccccc}
\hline
\toprule
 &
   &
   &
  \multicolumn{2}{c}{\textbf{Modification}} &
  \multicolumn{4}{c}{\textbf{Blending}} &
  \multicolumn{3}{c}{\textbf{Neuron Hijacking}} &
  \multicolumn{2}{c}{\textbf{Defense Adaptive }} \\ \cline{4-14} 
\multirow{-2}{*}{\textbf{Datasets}} &
  \multirow{-2}{*}{\textbf{Models}} &
  \multirow{-2}{*}{\textbf{Methods}} &
  BadNets &
  \begin{tabular}[c]{@{}c@{}}Dynamic\\ Backdoor\end{tabular} &
  \begin{tabular}[c]{@{}c@{}}Poison\\ frogs\end{tabular} &
  \begin{tabular}[c]{@{}c@{}}Hidden\\ Trigger\end{tabular} &
   BP &
   \begin{tabular}[c]{@{}c@{}}Sleeper\\ Agent\end{tabular} &
  \begin{tabular}[c]{@{}c@{}}TrojanNN \\ (Apple)\end{tabular} &
  \begin{tabular}[c]{@{}c@{}}TrojanNN \\ (Face)\end{tabular} &
  SIG &
  ABE &
  DFST \\ \hline
\multirow{27}{*}{MNIST}      & \multirow{9}{*}{LeNet-1}  & NC                       & 82.6\%             & 78.6\%              & N               & N               & 65.2\%          & 61.4\%          & 60.4\%           & 51.6\%          & 61.6\%          & N                     & /                    \\
                             &                           & TABOR                    & 83.2\%             & 79.2\%              & N               & N               & 62.4\%          & 68.0\%          & 70.6\%           & 69.2\%          & 65.2\%          & N                     & /                    \\
                             &                           & ABS                      & 89.8\%             & 81.2\%              & 68.0\%          & N               & 77.6\%          & 60.2\%          & 76.8\%           & 81.8\%          & 67.8\%          & N                     & /                    \\
                             &                           & TND                      & 88.8\%             & 82.2\%              & 65.2\%          & N               & 70.0\%          & 72.6\%          & 70.4\%           & 64.2\%          & 79.6\%          & N                     & /                    \\
                             &                           & K-Arm                    & \textbf{92.0\%}    & 84.0\%              & 78.4\%          & 62.8\%          & 78.2\% & 74.8\%          & 78.2\%           & 80.2\%          & 84.2\%          & 57.0\%                & /                    \\
                             &                           & ANP                      & \textbf{91.4\%}    & 82.6\%              & 78.2\%          & 58.8\%          & 63.2\% & 54.6\%          & 67.4\%           & 69.0\%          & 74.0\%          & 64.8\%                & /                    \\
                             &                           & TopoTrigger              & \textbf{90.8\%}    & 81.2\%              & 90.8\%          & 70.4\%          & \textbf{80.2\%} & 70.2\%          & 75.2\%           & 78.2\%          & 83.8\%          & 80.6\%                & /                    \\
                             &                           & UNICORN                  & \textbf{91.2\%}    & 83.4\%              & 92.1\%          & 69.0\%          & 77.0\% & 72.4\%          & 80.0\%           & 81.4\%          & 81.6\%          & 81.2\%                & /                    \\
                             &                           & CatchBackdoor            & 91.0\%             & \textbf{84.6\%}     & \textbf{94.0\%} & \textbf{72.6\%} & 77.8\%          & \textbf{78.0\%} & \textbf{81.6\%}  & \textbf{82.4\%} & 82.0\%          & \textbf{86.0\%}       & /                    \\ \cline{2-14} 
                             & \multirow{9}{*}{LeNet-4}  & NC                       & 82.8\%             & 80.0\%              & N               & N               & 72.0\%          & 64.2\%          & 63.0\%           & 55.4\%          & 69.4\%          & N                     & /                    \\
                             &                           & TABOR                    & 90.4\%             & 83.2\%              & N               & N               & 72.4\%          & 78.0\%          & 63.8\%           & 62.0\%          & 64.8\%          & N                     & /                    \\
                             &                           & ABS                      & \textbf{92.0\%}    & 83.4\%              & 79.0\%          & 58.8\%          & 84.2\%          & 82.4\%          & 79.2\%           & 70.2\%          & 79.0\%          & N                     & /                    \\
                             &                           & TND                      & 90.2\%             & 80.2\%              & 64.2\%          & N               & 78.6\%          & 82.0\%          & 64.4\%           & 65.0\%          & 72.4\%          & N                     & /                    \\
                             &                           & K-Arm                    & 91.0\%             & 78.6\%              & 82.6\%          & 58.8\%          & \textbf{85.0\%} & \textbf{86.8\%} & 80.0\%           & 71.0\%          & 78.0\%          & 59.8\%                & /                    \\
                             &                           & ANP                      & 86.2\%             & 79.8\%              & 74.2\%          & 47.8\%          & 73.4\%          & 76.2\%          & 59.8\%           & 61.6\%          & 73.4\%          & 53.8\%                & /                    \\
                             &                           & TopoTrigger              & 77.0\%             & 83.6\%              & 79.2\%          & 76.8\%          & 71.8\%          & 80.8\%          & 72.6\%           & 70.6\%          & 82.8\%          & \textbf{93.4\%}                & /                    \\
                             &                           & UNICORN                  & 92.8\%             & 75.8\%              & 71.6\%          & 65.6\%          & 86.0\%          & 73.2\%          & 77.8\%           & 65.8\%          & 86.2\%          & 79.8\%                & /                    \\
                             &                           & CatchBackdoor            & 91.2\%             & \textbf{83.8\%}     & \textbf{91.0\%} & \textbf{70.6\%} & 83.2\%          & 82.0\%          & \textbf{81.0\%}  & \textbf{73.6\%} & 85.4\%          & 87.0\%       & /                    \\ \cline{2-14} 
                             & \multirow{9}{*}{LeNet-5}  & NC                       & 80.6\%             & 78.4\%              & N               & N               & N               & 56.0\%          & 61.6\%           & 62.6\%          & 78.4\%          & N                     & /                    \\
                             &                           & TABOR                    & 89.2\%             & 80.0\%              & N               & N               & 66.4\%          & 64.6\%          & 64.4\%           & 59.4\%          & 72.0\%          & N                     & /                    \\
                             &                           & ABS                      & 91.8\%             & \textbf{83.6\%}     & 74.0\%          & 51.8\%          & 78.6\%          & 74.4\%          & 75.2\%           & 71.8\%          & 80.0\%          & N                     & /                    \\
                             &                           & TND                      & 90.7\%             & 81.4\%              & N               & N               & 72.6\%          & 71.0\%          & 68.6\%           & 64.8\%          & 80.6\%          & N                     & /                    \\
                             &                           & K-Arm                    & \textbf{93.4\%}    & 81.6\%              & 84.4\%          & 58.2\%          & 78.4\%          & \textbf{78.0\%} & 74.0\%           & \textbf{81.2\%} & 83.4\%          & N                     & /                    \\
                             &                           & ANP                      & 81.8\%             & 62.8\%              & 69.8\%          & 60.4\%          & 58.8\%          & 68.6\%          & 58.8\%           & 64.0\%          & 73.6\%          & N                & /                    \\
                             &                           & TopoTrigger              & 94.0\%             & 79.2\%              & 84.6\%          & 56.6\%          & 66.4\%          & 59.8\%          & 66.8\%           & 59.8\%          & 85.4\%          & 67.8\%                & /                    \\
                             &                           & UNICORN                  & 91.6\%             & 87.6\%              & 82.8\%          & 77.6\%          & 65.6\%          & 76.0\%          & 58.8\%           & 78.0\%          & 80.6\%          & 82.2\%                & /                    \\
                             &                           & CatchBackdoor            & 92.2\%             & 82.6\%              & \textbf{93.0\%} & \textbf{74.8\%} & \textbf{80.0\%} & \textbf{78.0\%} & \textbf{78.0\%}  & 80.2\%          & 82.4\%          & \textbf{91.0\%}       & /                    \\ \hline
\multirow{18}{*}{CIFAR-10}   & \multirow{9}{*}{AlexNet}  & NC                       & 81.6\%             & 74.6\%              & N               & N               & N               & N               & 51.2\%           & N               & 54.6\%          & N                     & N                    \\
                             &                           & TABOR                    & 87.6\%             & 76.6\%              & N               & N               & 56.0\%          & N               & 67.2\%           & 66.6\%          & 57.2\%          & N                     & N                    \\
                             &                           & ABS                      & 90.4\%             & 79.4\%              & 75.0\%          & N               & 68.0\%          & 72.6\%          & 75.8\%           & 64.2\%          & 51.0\%          & N                     & N                    \\
                             &                           & TND                      & 89.5\%             & 77.6\%              & 50.8\%          & 52.6\%          & 70.8\%          & 70.0\%          & 70.8\%           & 60.0\%          & 64.6\%          & N                     & N                    \\
                             &                           & K-Arm                    & 92.0\%             & 82.0\%              & 75.2\%          & 58.6\%          & \textbf{74.2\%} & 66.6\%          & 76.8\%           & 76.4\%          & 72.8\%          & N                     & 51.2\%               \\
                             &                           & ANP                      & 86.6\%             & 78.4\%              & 63.2\%          & 66.2\%          & 61.8\%          & N         & 66.2\%           & 54.6\%          & 74.0\%          & N                & N               \\
                             &                           & TopoTrigger              & 82.6\%             & 72.0\%              & 78.8\%          & 67.4\%          & 65.4\%          & 53.2\%          & 73.8\%           & 61.4\%          & 63.6\%          & 77.4\%                & 60.0\%               \\
                             &                           & UNICORN                  & 92.0\%             & 72.6\%              & 82.0\%          & 68.8\%          & 69.0\%          & 67.2\%          & \textbf{80.8}\%           & 63.6\%          & 79.6\%          & 79.8\%                & 63.2\%               \\
                             &                           & CatchBackdoor            & \textbf{93.6\%}    & \textbf{83.2\%}     & \textbf{87.0\%} & \textbf{70.8\%} & 72.4\%          & \textbf{70.2\%} & 79.4\%  & \textbf{79.2\%} & \textbf{80.8\%} & \textbf{83.0\%}       & \textbf{61.4\%}      \\ \cline{2-14} 
                             & \multirow{9}{*}{ResNet20} & NC                       & 83.6\%             & 78.8\%              & N               & N               & N               & N               & 60.0\%           & 53.8\%          & 64.6\%          & N                     & N                    \\
                             &                           & TABOR                    & 84.0\%             & 79.8\%              & N               & N               & 56.8\%          & N               & 63.0\%           & 50.2\%          & 60.6\%          & N                     & N                    \\
                             &                           & ABS                      & 89.6\%             & 82.8\%              & 77.0\%          & 53.2\%          & 72.0\%          & N               & 81.0\%           & 60.6\%          & 70.0\%          & N                     & 56.8\%               \\
                             &                           & TND                      & 88.3\%             & 78.2\%              & N               & N               & 68.8\%          & N               & 74.8\%           & 51.6\%          & 68.4\%          & N                     & N                    \\
                             &                           & K-Arm                    & 91.0\%             & 79.4\%              & 81.4\%          & 70.2\%          & 72.0\%          & 51.4\%          & 81.2\%           & 78.0\% & 75.2\%          & N                     & 56.8\%               \\
                             &                           & ANP                      & 80.2\%             & 73.4\%              & N          & N          & 60.0\%          & 33.4\%          & 75.4\%           & 52.4\%          & 54.0\%          & N                & N               \\
                             &                           & TopoTrigger              & 89.6\%             & 78.2\%              & 83.4\%          & 70.0\%          & 65.4\%          & 68.6\%          & 70.2\%           & 71.6\%          & 72.0\%          & 78.0\%                & 53.8\%               \\
                             &                           & UNICORN                  & 87.8\%             & 84.0\%              & 79.2\%          & 72.2\%          & 78.6\%          & 64.0\%          & 80.4\%           & \textbf{79.0\%}          & 75.2\%          & 81.6\%                & 57.2\%               \\
                             &                           & CatchBackdoor            & \textbf{90.8\%}    & \textbf{83.4\%}     & \textbf{88.0\%} & \textbf{74.0\%} & \textbf{80.0\%} & \textbf{70.6\%} & \textbf{82.2\%}  & 77.2\%          & \textbf{73.8\%} & \textbf{85.0\%}       & \textbf{60.8\%}      \\ \hline
\multirow{18}{*}{a-ImageNet} & \multirow{9}{*}{VGG-16}   & NC                       & 81.4\%             & 76.2\%              & N               & N               & N               & N               & 57.2\%           & 51.2\%          & 54.2\%          & N                     & N                    \\
                             &                           & TABOR                    & 82.8\%             & 76.8\%              & N               & N               & N               & N               & 64.8\%           & 66.2\%          & 63.8\%          & N                     & N                    \\
                             &                           & ABS                      & 86.0\%             & \textbf{79.8\%}     & 69.0\%          & N               & 56.2\%          & 78.4\%          & 73.0\%           & 68.8\%          & 69.0\%          & N                     & 51.6\%               \\
                             &                           & TND                      & 86.1\%             & 77.2\%              & 58.0\%          & N               & N               & 67.0\%          & 72.6\%           & 65.0\%          & 77.4\%          & N                     & N                    \\
                             &                           & K-Arm                    & 89.0\%             & 78.8\%              & 69.2\%          & N               & 70.8\%          & 78.4\% & 76.8\%           & \textbf{77.6\%} & 87.2\%          & N                     & N                    \\
                             &                           & ANP                      & 80.2\%             & 69.2\%              & 56.0\%          & 63.4\%          & 45.6\%          & 73.6\%          & 72.8\%           & 59.4\%          & 64.6\%          & N                & N               \\
                             &                           & TopoTrigger              & 81.8\%             & 70.4\%              & 72.4\%          & 68.8\%          & 70.8\%          & \textbf{80.2\%}          & 78.4\%           & 66.2\%          & 82.4\%          & 57.8\%                & 60.2\%               \\
                             &                           & UNICORN                  & 89.4\%             & 72.4\%              & 59.0\%          & 70.8\%          & 71.8\%          & 60.0\%          & 75.6\%           & 69.0\%          & 84.8\%          & 53.4\%                & 51.4\%               \\
                             &                           & CatchBackdoor            & \textbf{90.2\%}    & 79.4\%              & \textbf{76.0\%} & \textbf{76.4\%} & \textbf{71.4\%} & 77.8\%          & \textbf{80.2\%}  & \textbf{77.6\%} & \textbf{88.4\%} & \textbf{79.0\%}       & \textbf{54.8\%}      \\ \cline{2-14} 
                             & \multirow{9}{*}{VGG-19}   & NC                       & 79.2\%             & 76.4\%              & N               & N               & N               & N               & 56.4\%           & 57.2\%          & 56.8\%          & N                     & N                    \\
                             &                           & TABOR                    & 80.2\%             & 78.4\%              & N               & N               & N               & 53.2\%          & 58.8\%           & 63.2\%          & 68.2\%          & N                     & N                    \\
                             &                           & ABS                      & 85.6\%             & 77.8\%              & 65.0\%          & N               & 62.0\%          & 77.0\%          & 74.4\%           & 70.2\%          & 72.4\%          & N                     & N                    \\
                             &                           & TND                      & 84.3\%             & 77.0\%              & 57.0\%          & N               & 58.4\%          & 77.0\%          & 70.2\%           & 53.4\%          & 70.0\%          & N                     & N                    \\
                             &                           & K-Arm                    & 88.8\%             & 78.2\%              & 70.0\%          & 50.8\%          & 64.2\%          & 74.0\%          & \textbf{79.4\%}  & \textbf{75.2\%} & 83.4\%          & N                     & 50.0\%               \\
                             &                           & ANP                      & 84.4\%             & 76.2\%              & N          & N          & 56.8\%          & 57.6\%          & N  & N & 67.2\%          & N               & N               \\
                             &                           & TopoTrigger              & 89.2\%             & 77.6\%              & 73.2\%          & 56.4\%          & 72.2\%          & 69.2\%          & 70.6\%  & 72.2\% & 83.8\%          & N                & 51.2\%               \\
                             &                           & UNICORN                  & 91.6\%             & 79.4\%              & 76.0\%          & 52.0\%          & 56.6\%          & 70.6\%          & 76.0\%  & 70.6\% & 88.6\%          & 68.4\%                & 55.8\%               \\
                             &                           & CatchBackdoor            & \textbf{90.0\%}    & \textbf{80.8\%}     & \textbf{79.0\%} & \textbf{76.8\%} & \textbf{80.6\%} & \textbf{77.2\%} & 79.2\%           & 73.2\%          & \textbf{84.0\%} & \textbf{75.0\%}       & \textbf{56.6\%}      \\ \bottomrule
\end{tabular}
}
\label{LCR Comparison}
\end{table}

For clean label attacks like BP and Sleeper Agent, CatchBackdoor shows inferior performance than K-Arm, especially on LeNet-4 and AlexNet. We suppose that critical neural path is not strongly correlated with the clean trojaned label. So the finally-converged critical trojan path will be misleading to trigger trojaned behaviors. We will leave improvements on it in the future work. 
As for defense adaptive attacks that try to evade possible defenses, the superiority of CatchBackdoor can be obviously observed. 

We also notice that LCR decreases on large datasets and complex models. For instance, on VGG19 of a-ImageNet, LCR is around 76\% on average. The reason lies in that the increasing depth of the model may lead to the redundancy of neurons. Some neuron activation values in the neural path may not be significantly larger than that of redundant neurons, which leads to the decrease of LCR for those models.
We have compared the activation value and frequency between benign, trojaned, and reversed examples for interpretation.

We further extend the effectiveness of Catchbackdoor to transformer-based models such as ViT~\cite{dosovitskiy2020image} and Deit~\cite{touvron2021training}, as well as against advanced attacks~\cite{nguyen2021wanet} and large-scale datasets. Additionally, we demonstrate the efficacy of Catchbackdoor at low trojan rates. Detailed experimental results and visualizations are provided in the \textbf{supplementary materials}.

\subsection {Detection Extensibility \label{varying scene}}
We conduct detection when applying different trigger sizes, when the training data is unavailable, and when using adaptive attacks. We have also tested pre-trained models on Caffe Model Zoo~\cite{modelzoo} and provided the results in the \textbf{supplementary materials}.

\begin{wraptable}{r}{0.4\linewidth}
\caption{Sensitivity on trigger size.}
\label{trigger_sen}
\centering
\resizebox{1\linewidth}{!}{ 
\begin{tabular}{cccccc}
\toprule
\multirow{2}{*}{\textbf{Datasets}} &
  \multirow{2}{*}{\textbf{Models}} &
  \multirow{2}{*}{\textbf{\begin{tabular}[c]{@{}c@{}}Trigger\\ Size\end{tabular}}} &
  \multicolumn{3}{c}{\textbf{LCR}} \\ \cline{4-6} 
                          &                          &       & NC      & K-Arm   & CatchBackdoor \\ \hline
\multirow{6}{*}{MNIST}    & \multirow{6}{*}{LeNet-5} & 3$\times$3   & 80.0\% & 89.6\% & 87.4\%       \\
                          &                          & 7$\times$7   & 63.2\% & 74.4\% & 78.4\%       \\
                          &                          & 10$\times$10 & N       & 68.0\% & 65.2\%       \\
                          &                          & 13$\times$13 & N       & N       & 60.2\%       \\
                          &                          & 18$\times$18 & N       & N       & 57.8\%       \\
                          &                          & 22$\times$22 & N       & N       & 56.2\%       \\ \hline
\multirow{6}{*}{CIFAR-10} & \multirow{6}{*}{AlexNet} & 3$\times$3   & 81.6\% & 92.0\% & 93.6\%       \\
                          &                          & 7$\times$7   & 74.2\% & 86.6\% & 86.6\%       \\
                          &                          & 10$\times$10 & N       & 78.4\% & 80.4\%       \\
                          &                          & 13$\times$13 & N       & 66.0\% & 74.2\%       \\
                          &                          & 18$\times$18 & N       & N       & 72.8\%       \\
                          &                          & 22$\times$22 & N       & N       & 70.2\%       \\ \bottomrule
\end{tabular}
}
\end{wraptable}
\subsubsection{Detection Sensitivity to Trigger Size}

We apply BadNets on LeNet-5 of MNIST and AlexNet of CIFAR-10 with a fixed poisoning rate 5\%. For the trigger, we gradually increase its size from 3$\times$3 to 22$\times$22. We generate 500 examples to calculate LCR. Results are shown in Table \ref{trigger_sen}. We can observe that CatchBackdoor is more robust to trigger size than NC and K-Arm. For instance, when trigger size is larger than 13$\times$13, LCR of NC and ABS is lower than 50\%, i.e., these trojaned models cannot be detected. CatchBackdoor can still achieve detection when even half of the image is covered by the trojaned triggers. This is because trojan path is independent of trigger size, consistent with our assumption.

\begin{figure*}[htbp]
\centering
\begin{minipage}{0.45\linewidth}
\centering
    \captionof{table}{LCR by generating examples with random noise.}
\resizebox{\linewidth}{!}{ 
\begin{tabular}{cccccc}
\toprule
\multirow{2}{*}{\textbf{Datasets}} &
  \multirow{2}{*}{\textbf{Models}} &
  \multicolumn{4}{c}{\textbf{Attacks}} \\ \cline{3-6} 
 &
   &
  BadNets &
  \begin{tabular}[c]{@{}c@{}}Poison\\ frogs\end{tabular} &
  \begin{tabular}[c]{@{}c@{}}TrojanNN\\ (Face)\end{tabular} &
  ABE \\ \hline
MNIST         & LeNet-5  & 90.2\% & 81.8\% & 64.0\% & 71.4\% \\
CIFAR-10      & ResNet20 & 83.6\% & 78.6\% & 63.2\% & 74.4\% \\
a-ImageNet & VGG19    & 84.8\% & 60.4\% & 70.0\% & 64.8\% \\ \bottomrule
\end{tabular}}
\label{without input}
\end{minipage}
\hfill
\begin{minipage}{0.5\linewidth}
\centering
\caption{Generated testing examples by CatchBackdoor.} 
\label{trigger_sen_fig} 
\subfloat[plane]{\includegraphics[width=0.23\linewidth]{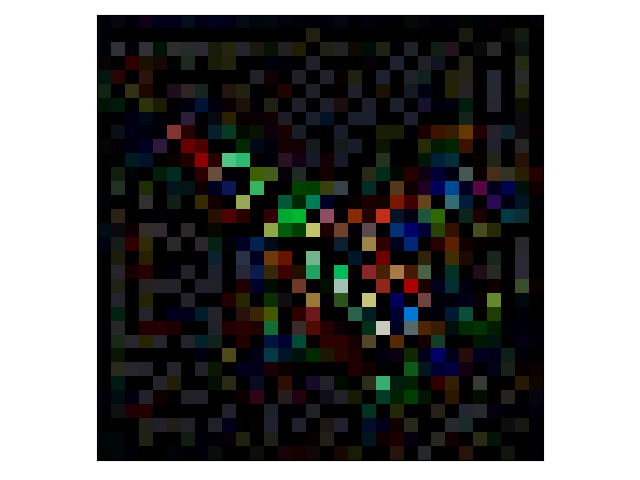}} 
\subfloat[bird]{\includegraphics[width=0.23\linewidth]{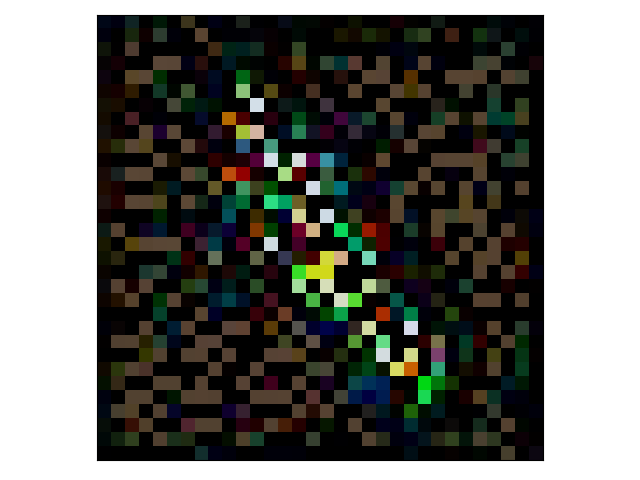}}
\subfloat[cat]{\includegraphics[width=0.23\linewidth]{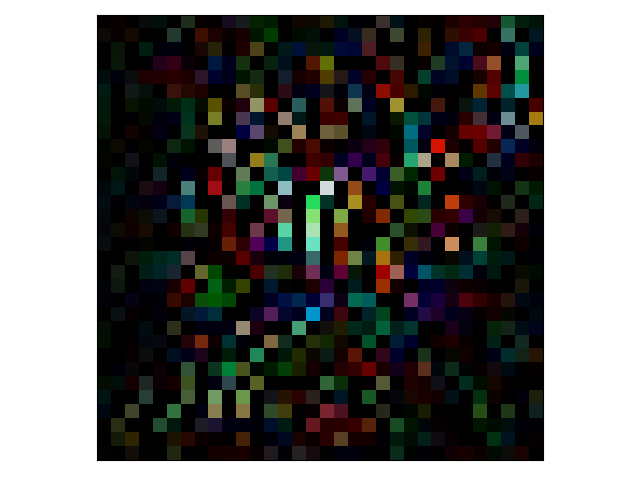}}
\subfloat[deer]{\includegraphics[width=0.23\linewidth]{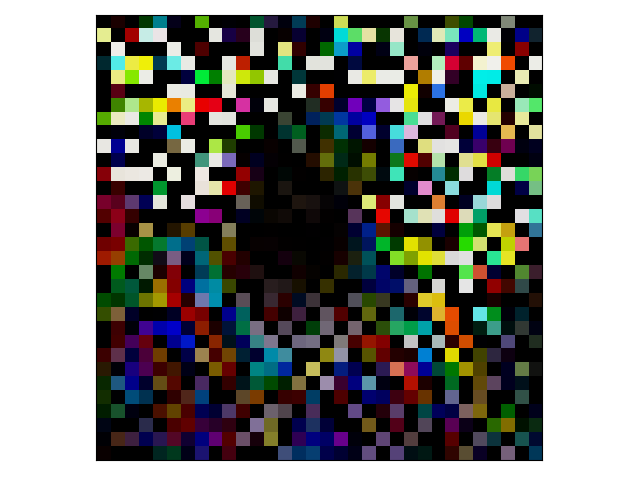}}
\end{minipage}
\end{figure*}

\subsubsection{Detection Without Training Data\label{test withou data}} 
We conduct experiments on LeNet-5 of MNIST, ResNet20 of CIFAR-10 and VGG19 of a-ImageNet. For each model, we generate 500 reversed examples from random noise. Detection performance is measured by LCR shown in Table \ref{without input}. 
Reversed examples generated on ResNet20 of CIFAR-10 under Poison frogs are shown in Fig.~\ref{trigger_sen_fig}, where the reversed trigger is added to the black image (the pixel value is set to 0) for better visualization. Trojaned labels are attached in the corresponding caption. 
From the table, when starting with random noise, LCR of CatchBackdoor is still high, around 73\% on average. Even on advanced attack ABE, no significant decrease can be seen, compared with that in Table \ref{LCR Comparison}. CatchBackdoor can conduct testing example generation leveraging on the diversity of input examples from random noise, which is different from example-dependent testing. Setting the threshold of LCR at 50\%, most trojaned models can still be detected. CatchBackdoor still performs well mainly because trojan path is not directly related to benign seeds, and different benign paths all converge to one trojan path. Thus, the random noise can be adopted to construct a benign path. More visualizations of critical trojan path fuzzed from random noise are in the \textbf{supplementary materials}.

\subsubsection{Detection against Adaptive Attacks}
\begin{figure*}[htbp]
\centering
\begin{minipage}[c]{0.45\linewidth}
\centering
\small
\captionof{table}{Detection results of adaptive attacks.}
\label{Adaptive}
\resizebox{\linewidth}{!}{ 
\begin{tabular}{cccccc}
\toprule
\multirow{2}{*}{\textbf{Datasets}} & \multirow{2}{*}{\textbf{Models}} & \multicolumn{2}{c}{\textbf{S-AA}} & \multicolumn{2}{c}{\textbf{I-AA}} \\ \cline{3-6} 
                                   &                                  & \textbf{ASR}    & \textbf{LCR}    & \textbf{ASR}    & \textbf{LCR}    \\ \hline
\multirow{2}{*}{MNIST}             & LeNet-4                          & 92.0\%         & 88.3\%         & 91.8\%         & 87.5\%         \\
                                   & LeNet-5                          & 91.6\%         & 86.4\%         & 90.4\%         & 85.1\%         \\ \hline
\multirow{2}{*}{CIFAR-10}          & AlexNet                          & 88.3\%         & 85.3\%         & 88.5\%         & 85.0\%         \\
                                   & ResNet20                         & 87.8\%         & 85.7\%         & 88.4\%         & 86.1\%         \\ \hline
\multirow{2}{*}{ImageNet}          & VGG16                            & 84.6\%         & 85.1\%         & 86.5\%         & 85.3\%         \\
                                   & VGG19                            & 85.5\%         & 85.4\%         & 86.1\%         & 85.9\%         \\ \bottomrule
\end{tabular}
}
\end{minipage}
\hfill
\begin{minipage}[c]{0.5\linewidth}
\centering
\caption{LCR under different $k_1$ and $k_2$}
\centering
\subfloat[Influence of $k_1$]{%
    \includegraphics[width=0.47\linewidth]{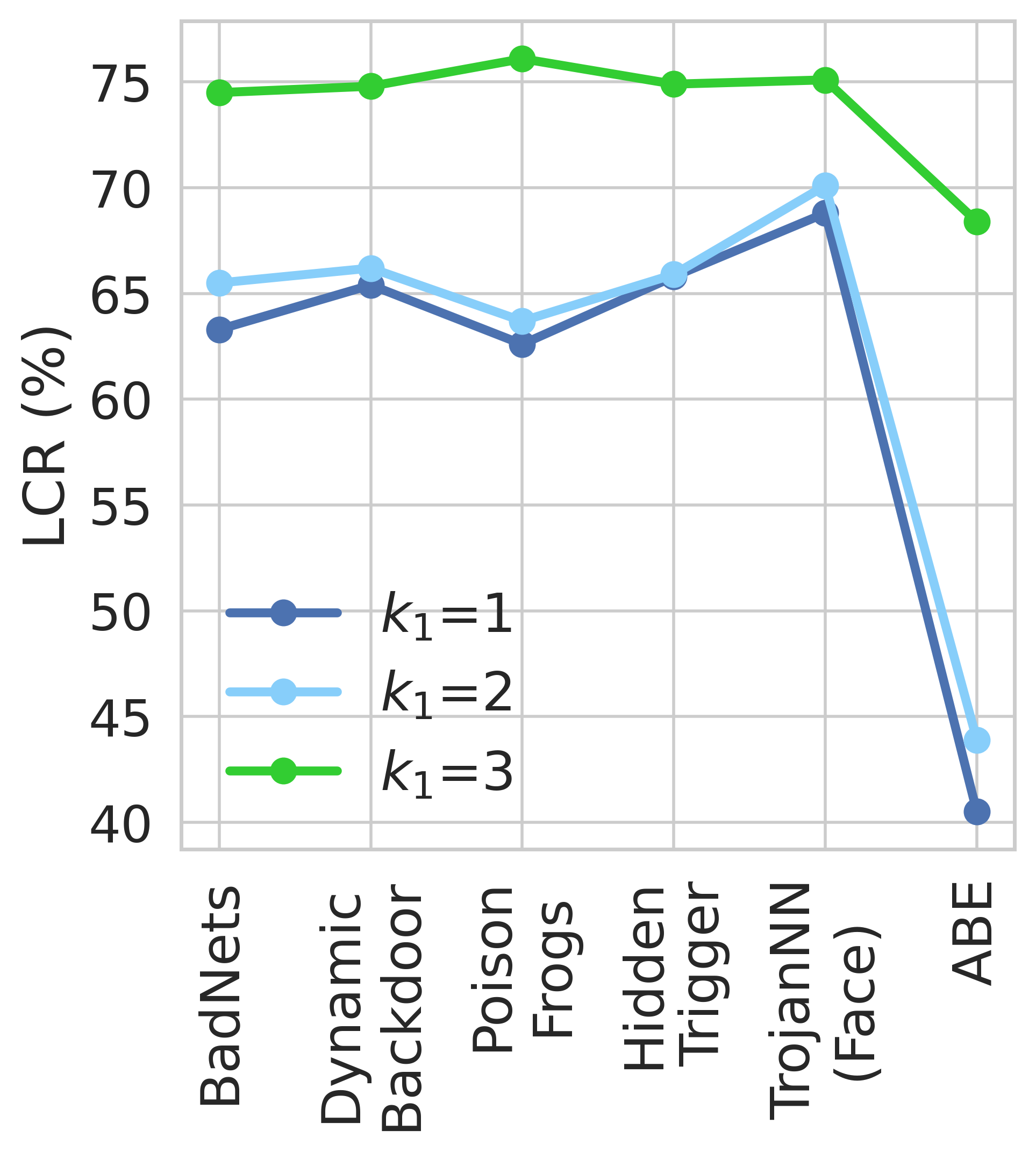}%
}
\subfloat[Influence of $k_2$]{%
    \includegraphics[width=0.47\linewidth]{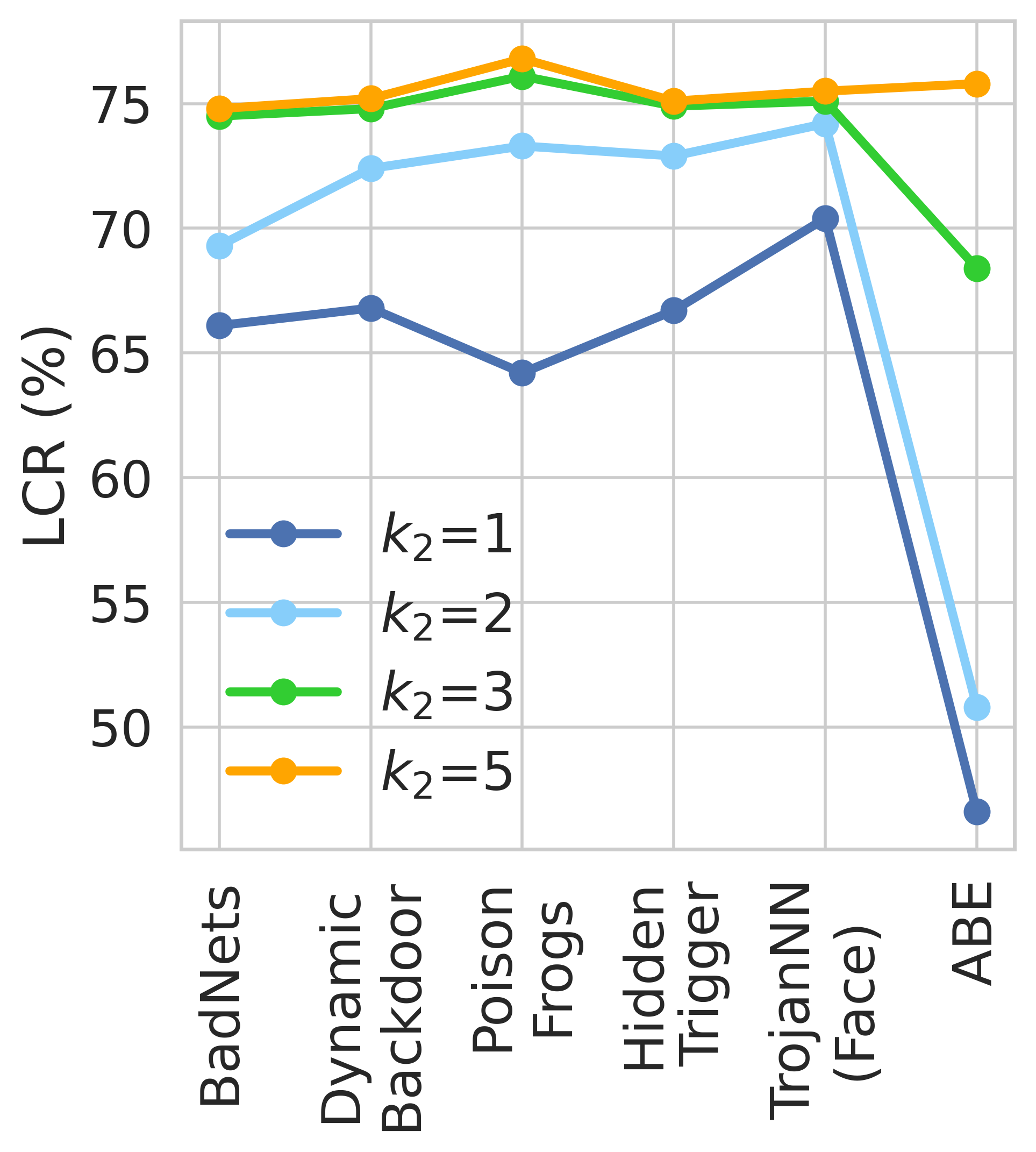}%
}
\label{fig parameter k}
\end{minipage}
\end{figure*}

We evaluate the CatchBackdoor's detection performance under potential countermeasures where the attacker knows our detection in advance and tries to bypass it. We design two types of adaptive trojan attacks, i.e., standardized adaptive attack (S-AA) and imitation adaptive attack (I-AA) and further evaluate CatchBackdoor on them. Details of these two attacks can be seen the \textbf{supplementary materials}. Results of S-AA and I-AA are shown in Table~\ref{Adaptive}. 
 
We observe that while the adaptive attacks do increase the difficulty, the detection performance of our method is still remarkable under adaptive settings, with LCR over 85\%. This indicates that backdoors crafted by adaptive attacks can still be hunted by CatchBackdoor. It is harder to identify critical neurons when the difference in neuron contribution is smaller. On this occasion, neuron frequency plays an important role. By selecting neurons based on activated frequency, neurons responsible for trojaned behaviors can still be identified and targeted through neural path fuzzing. Afterward, errors can be triggered by CatchBackdoor.

\subsection{Parameter Sensitivity Analysis \label{k1k2}}
We analyze the influence of $k_1$ and $k_2$, which control the path selection of the convolution layers and fully connected layers, respectively.

We measure the LCR of 500 testing examples generated under different $k_1$ and $k_2$ on VGG19 of a-ImageNet. For BadNets, MP is used with 7$\times$7 patch size. Trojan rate for all attacks is 0.1. The results are shown in Fig.~\ref{fig parameter k}, where abscissa denotes different trojan attacks.

As observed, on most cases, LCR all exceeds 50\%. This indicates that various backdoors can still be detected under different $k_1$ and $k_2$. So we can conclude that the detection capability of CatchBackdoor is stable under different $k_1$ and $k_2$ but it will decrease when faced with defense adaptive attacks as ABE. It is easy to understand that by controlling more neurons that contribute to the model's predictions, more errors can be triggered. So the increase of $k_1$ and $k_2$ can effectively raise the value of LCR. 

For defense adaptive attack ABE, we find that the testing examples generated by only a single neuron cannot achieve high LCR. This is consistent with our assumption that trojaned behaviors are produced by the joint action of multiple neurons. So when $k_1$ and $k_2$ are both set to 3, LCR of ABE is over 65\%, which shows backdoors carfted by ABE can be hunted by CatchBackdoor.

Besides, a larger value of $k$ is not related to a better effect. With the increase of $k_1$ and $k_2$, neural path selection will be more time-consuming, for introducing too many redundant neurons.

\section{Conclusion}
This paper proposes the concept of neural path and we empirically find that trojaned behaviors are attributed to the trojan path, i.e., a neural path consisting of neurons that play decisive roles in changing model predictions. Motivated by it, we develop a backdoor detection method CatchBackdoor. We fuzz the neural path from the benign path, to generate reversed examples that can trigger errors due to backdoors. Extensive experiments have verified the effectiveness of CatchBackdoor in hunting various backdoors. Besides, our method is independent of trigger size and can perform detection without benign training data.

\section*{Acknowledgements}
We extend our gratitude to all authors, reviewers, and the chair for their invaluable contributions. Additionally, we would like to express our appreciation to Dongping Chen for providing computational resources.


\newpage

\title{Supplementary Material of\\ CatchBackdoor: Backdoor Detection via Critical Trojan Neural Path Fuzzing} 

\titlerunning{CatchBackdoor}

\author{Haibo Jin\inst{1}\orcidlink{0000-0002-7244-7659} \and
Ruoxi Chen\inst{2}\orcidlink{0000-0003-2626-5448} \and
Jinyin Chen\inst{3, 2}\and
Haibin Zheng\inst{3, 2} \and\\
Yang Zhang\inst{1} \and
Haohan Wang\inst{1}\orcidlink{0000-0002-1826-4069}
}

\authorrunning{H.~Jin et al.}

\institute{School of Information Sciences, University of Illinois Urbana-Champaign \email{\{haibo,yzhangnd,haohanw\}@illinois.edu}\\ \and
College of Information Engineering, Zhejiang University of Technology \\
\email{\{2112003149, chenjinyin\}@zjut.edu.cn}\\ \and
Institute of Cyberspace Security, Zhejiang University of Technology\\
\email{haibinzheng320@gmail.com}}

\maketitle

In this supplementary document, we provide additional experiments and settings for CatchBackdoor. We first present the pseudo-code in Section~\ref{alg}. Next, we describe the experimental details in Section~\ref{details}. A running example is provided in Section~\ref{running_example}. More experimental results of the effectiveness of CatchBackdoor are shown in Section~\ref{effc}, followed by a detailed investigation of neuron behaviors in Section~\ref{neuron_behaviors}. An intuitive understanding of CatchBackdoor is given in both the input feature space and the high-dimensional representative feature space in hidden layers in Section~\ref{Visualizations}. Additionally, The detection results for the online model zoo are presented in Section~\ref{model_zoo_res}, and experiments against possible adaptive attacks are detailed in Section~\ref{Adaptive_Attacks}. We also use images generated by CatchBackdoor to repair the model in Section~\ref{robust improve}. Time complexity is discussed in Section~\ref{time}. Finally, we evaluate the effect of random noise triggers on the benign path in Section~\ref{appendix_noise} and visualize the top-5 neural paths in Section~\ref{Visualizationtop-5}.

\section{CatchBackdoor Algorithm}\label{alg}

\subsection{Benign Path Construction}
Constructing a benign path is crucial for ensuring network integrity. This process involves identifying and linking key neurons that significantly impact the network's output, while excluding neurons from the fully connected layer to maintain path diversity. The pseudocode for constructing a benign path is outlined in Algorithm~\ref{Benign_Path}.

\begin{algorithm}[H]
\small
\caption{Benign Path Construction}
\begin{algorithmic}[1]
\State \textbf{Input:} Model parameters, benign seed \( x \), number of neurons \( k \)
\State \textbf{Output:} Benign Path $\Psi_{b} (x)$
\State Calculate neuron contributions using \textbf{Eq. 1}.
\State Arrange neurons in descending order of contribution.
\State Select top-\( k \) neurons as benign critical neurons.
\State Form $\Psi_{b} (x)$ using \textbf{Eq. 2}
\State \textbf{return} $\Psi_{b} (x)$
\end{algorithmic}
\label{Benign_Path}
\end{algorithm}

\subsection{Critical Trojan Path Identification Through Fuzzing}
We conduct path fuzzing to identify the critical trojan path, from which triggers will be reversed afterward. The process involves maximizing neuron contributions within the benign path and calculating the activation frequency of each neuron from the aggregation of fuzzed paths. By linking critical trojan neurons with data flow, we form the critical trojan path that can trigger the correct trojaned label, similar to trojaned examples. Algorithm~\ref{critical_trojan_path} outlines the steps involved in this identification process.

\begin{algorithm}[H]
\small
\centering
\caption{Identification of Critical Trojan Path}
\label{critical_trojan_path}
\begin{algorithmic}[1]
\State \textbf{Input:} Model parameters, benign image $x$, iterations $S$, parameters $k$, frequency function $f(\cdot)$.
\State \textbf{Output:} Critical Trojan Path $\Psi (x)$
\State Initialize Fuzzed Path Set $H(x) = \{\}$
\State Initialize Fuzzed Path $\Psi_{f} (x)_1 = \Psi_{b} (x)$
\State Initialize Critical Trojan Neurons $\Omega (x)= \{\}$
\State Initialize Critical Trojan Path $\Psi_{cr} (x)= \{\}$
\For{$i = 2$ to $S$}
    \State Calculate $\xi_{n_{i,j}}(x)$ using \textbf{Eq. 1}
    \State Maximize $\xi_{n_{i,j}}(x), \forall n_{i,j} \in \Psi_{f} (x)_{i-1}$
    \State $H \leftarrow H \cup \Psi_{f}(x)_{i}$
\EndFor
\State $\Omega (x)=\bigcup_{i=1}^{l-1}\bigcup_{j=1}^{k} \arg\max_{n_{i,j}\in H} f(n_{i,j},x)$
\State Form $\Psi_{cr} (x)$ using \textbf{Eq. 2}
\State \Return $\Psi_{cr} (x)$
\end{algorithmic}
\end{algorithm}

\section{Detailed Experiment Setup}\label{details}
\subsection{Dataset and Model Configurations}
The dataset and model configurations are shown in Table~\ref{data_model}.

\begin{table}[htbp]
\huge
\centering
\vspace{-0.3cm}
\caption{Dataset and model configurations.}
\vspace{-0.3cm}
\resizebox{0.68\linewidth}{!}{
\begin{tabular}{cccclcc}
\hline
\textbf{Datasets}                   & \textbf{\#Categories}            & \textbf{\#Training Data}                              & \textbf{Models}     & \textbf{\#Params}      & \textbf{acc} \\ \hline
\multirow{3}{*}{MNIST}    & \multirow{3}{*}{10} & \multirow{3}{*}{50,000}      & LeNet-1   & 7,206       & 95.11\%  \\
                                              &                            &                            & LeNet-4   & 69,362      & 98.50\%  \\
                          &                     &                                                       & LeNet-5   & 107,786     & 98.90\%  \\ \hline
\multirow{2}{*}{CIFAR-10} & \multirow{2}{*}{10} & \multirow{2}{*}{50,000}      & AlexNet   & 87,650,186  & 91.24\%  \\
                                              &                            &                            & ResNet20 & 273,066      &93.04\%  \\ \hline
\multirow{2}{*}{a-ImageNet} & \multirow{2}{*}{10} & \multirow{2}{*}{13,000} &   VGG16    & 134,326,366 & 95.13\%  \\
                          &                     &                            &                             VGG19    & 139,638,622 & 95.40\%  \\ \hline
\end{tabular}
}
\label{data_model}
\vspace{-20pt}
\end{table}

\subsection{Trojan Attacks and Model Setting}
We use 11 attacks for evaluation, including modification attacks, blending attacks, neuron hijacking attacks, and defense adaptive attacks. Modification attacks include BadNets~\cite{DBLP:journals/corr/abs-1708-06733} and Dynamic Backdoor~\cite{SalemWBMZ22}. The patch size is 1$\times$1 on MNIST, 3$\times$3 on CIFAR-10, and 7$\times$7 on a-ImageNet. For BadNets, we randomly choose one kind of three patches to generate trojaned models. Blending attacks contain Poison frogs~\cite{shafahi2018poison}, Hidden trigger~\cite{saha2020hidden}, BullseyePolytope (BP)~\cite{aghakhani2021bullseye} and Sleeper Agent~\cite{souri2021sleeper}. As for neuron hijacking attacks, we adopted TrojanNN~\cite{DBLP:conf/ndss/LiuMALZW018}, including face stamp and apple stamp, and neuron frequency-based attack, SIG~\cite{barni2019new}. Two defense adaptive attacks include adversarial backdoor embedding (ABE)~\cite{tan2019bypassing}, and deep feature space trojan attack (DFST)~\cite{cheng2020deep}. DFST can not handle gray-scale images so it is not conducted on MNIST dataset. Since CatchBackdoor does not involve the training stage to mitigate the backdoor, so we do not take training stage defense methods.

In the experiments of Section 6.2, we train 50 benign and 50 trojaned models. They are trojaned by BadNets, Hidden trigger, and TrojanNN with trojan rates ranging from 0.1 to 0.3 (i.e., 0.1, 0.15, 0.2, 0.25, and 0.3), respectively.
In Section 6.3, For each dataset, we calculate the LCR of trojaned label on 50 benign and 50 trojaned models. They are trojaned by BadNets, Hidden trigger, and TrojanNN with trojan rates ranging from 0.1 to 0.3 (i.e., 0.1, 0.15, 0.2, 0.25, and 0.3), respectively. ASR of all trojaned models is over 90\%.

\section{Neural Path Fuzzing: A Running Example}\label{running_example}
 \begin{figure}[t]
\centering
\includegraphics[width=0.8\linewidth]{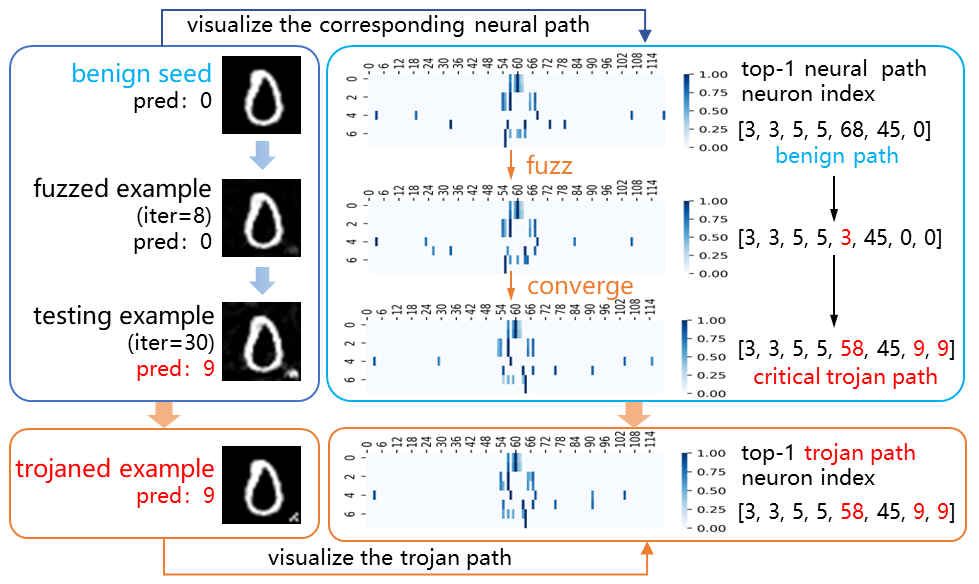}
\caption{Illustration of benign path, critical trojan path, and trojan path. }
\label{runningcase}
\end{figure}
We have illustrated the process of fuzzing neural path in Fig.~\ref{runningcase} on LeNet-5 of MNIST trojaned by BadNets. We set trojaned label 9, trojan rate=0.1, $batch\_ size=32$, $epoch=10$. ``iter'' refers to the current iteration number. We set the maximum iteration number $S$ to 30. In the heatmap, the contribution of neurons in the top-5 path after normalization to [0,1] is presented. The color goes deeper as the value grows. The x-axis is the neuron index while the y-axis is the layer index. The change of neuron index in the top-1 neural path is highlighted in red. 

To begin with, the benign seed (classified as ``0'') is input to the DNN. In the 8th iteration, the top-1 neural path changes in the fifth layer, i.e., neuron index from 68 to 3. But the predicted label remains unchanged. After 30 iterations, the path converges to the critical trojan path, whose neuron index of the top-1 path is the same as that of the trojan path. As a result, the testing example is misclassified as ``9'', i.e., the trojaned label.

After 30 rounds of iterations, the testing example show similar pattern with the same top-1 neural path, as the real trojaned example. This is consistent with our assumption that the converged critical trojan path plays a similar role as real trojan path, i.e., triggering trojaned behaviors. Besides, generated from the critical trojan path, reversed examples can serve as real triggers. Therefore, we leverage critical trojan path to trigger errors in trojaned models. Potential backdoors can be further hunted by identifying these errors. 

\section{Effectiveness of CatchBackdoor}\label{effc}
\subsection{All-to-All attacks}
We add detection experiments on attacks and show results in Table~\ref{all-to-all}. CatchBackdoor achieves a higher LCR on all-to-all attacks for identifying trojan paths for each trojan class.

\begin{table}[htbp]
\huge
\centering
\caption{Detection performance on all-to-all attacks}
\resizebox{0.85\linewidth}{!}{
\begin{tabular}{ccccccc}
\toprule
\textbf{Dataset} & \textbf{Models} & \textbf{Methods} & \textbf{BadNets} & \textbf{Hidden Trigger} & \textbf{SIG} & \textbf{ABE} \\ \hline
\multirow{4}{*}{CIFAR-10} & \multirow{4}{*}{AlexNet} & ANP & 80.2\% & 54.6\% & 71.2\% & N \\
& & TopoTrigger & 84.8\% & 72.2\% & 68.4\% & 62.4\% \\
& & UNICORN & 88.6\% & \textbf{76.8\%} & 75.6\% & 59.6\% \\
& & CatchBackdoor & \textbf{89.2\%} & 74.6\% & \textbf{79.0\%} & \textbf{65.8\%} \\ \hline
\multirow{4}{*}{a-ImageNet} & \multirow{4}{*}{VGG16} & ANP & 84.8\% & 67.4\% & 69.2\% & N \\
& & TopoTrigger & 86.0\% & 73.6\% & 80.8\% & 59.4\% \\
& & UNICORN & \textbf{87.2\%} & 73.0\% & \textbf{82.4\%} & 58.2\% \\
& & CatchBackdoor & 85.6\% & \textbf{74.6\%} & 79.6\% & \textbf{62.0\%} \\ \bottomrule
\end{tabular}
}
\label{all-to-all}
\end{table}
\subsection{Effectiveness on transformer-based models}
We attack ViT~\cite{dosovitskiy2020image} and DeiT~\cite{touvron2021training} on CIFAR-10 and a-ImageNet with BadNets, Hidden Trigger and WaNet~\cite{nguyen2021wanet} (all trojan rate=0.01). We selected a trojan rate of 0.01, with the benign label ``0'' converted to the trojan label ``9''. 
For BadNets, we achieved an attack success rate of 98.12\% on ViT and 97.26\%. For the Hidden Trigger attack, we achieved a success rate of 92.60\% on ViT and 91.74\% on DeiT. Under the WaNet attack, the success rates were 98.55\% on ViT and 96.42\% on DeiT. We further evaluated the detection performance using 500 benign examples, comparing our results with baselines. Table~\ref{transformer} shows CatchBackdoor maintains similar effectiveness on transformer-based models across different attacks.
\begin{table}[htbp]
\huge
\centering
\caption{Detection results on transformer-based models}
\resizebox{0.9\linewidth}{!}{
\begin{tabular}{ccccccc}
\toprule
\textbf{Dataset} & \textbf{Models} & \textbf{Attacks} & \textbf{ANP} & \textbf{TopoTrigger} & \textbf{UNICORN} & \textbf{CatchBackdoor} \\ \hline
\multirow{6}{*}{CIFAR-10} & \multirow{3}{*}{ViT} & BadNets & 78.2\% & 87.6\% & \textbf{90.4\%} & 90.2\% \\
& & Hidden Trigger & 69.6\% & 74.4\% & 78.2\% & \textbf{78.8\%} \\
& & WaNet & 62.0\% & 76.8\% & 77.6\% & \textbf{79.2\%} \\ \cline{2-7}
& \multirow{3}{*}{DeiT} & BadNets & 75.2\% & 86.4\% & \textbf{90.2\%} & 89.6\% \\
& & Hidden Trigger & 68.6\% & 78.8\%& 76.0\% & \textbf{79.6\%}  \\
& & WaNet & 64.8\% & \textbf{77.0\%} & 76.2\% & \textbf{77.0\%} \\ \hline
\multirow{6}{*}{a-ImageNet} & \multirow{3}{*}{ViT} & BadNets & 80.2\% & \textbf{90.2\%} & 89.2\% & 88.6\% \\
& & Hidden Trigger & 70.6\% & 75.6\% & 72.2\% & \textbf{77.2\%} \\
& & WaNet & 64.8\% & 74.0\% & \textbf{79.8\%} & 74.2\% \\ \cline{2-7}
& \multirow{3}{*}{DeiT} & BadNets & 76.8\% & 84.2\% & \textbf{91.6\%} & 90.8\% \\
& & Hidden Trigger & 67.4\% & 77.8\% & 77.8\% & \textbf{79.4\%} \\
& & WaNet & 63.0\% & 72.6\% & 78.0\% & \textbf{78.2\%} \\ \bottomrule
\end{tabular}
}
\label{transformer}
\end{table}

\subsection{Effectiveness at low traojan rates}
We run experiments with trojan rate=0.05 and 0.01 on VGG16 of a-ImageNet (Table~\ref{Rate}). CatchBackdoor achieves over 70\% LCR across different attacks, indicating it is not sensitive to the trojan rate.
\begin{table}[htbp]
\huge
\centering
\caption{Detection results on LCR with trojan rate=0.05 and 0.01}

\resizebox{0.7\linewidth}{!}{
\begin{tabular}{ccccccc}
\toprule
\textbf{\begin{tabular}[c]{@{}c@{}}Trojan\\ rate\end{tabular}} & \textbf{Methods} & \textbf{\begin{tabular}[c]{@{}c@{}}Poison\\ frogs\end{tabular}} & \textbf{\begin{tabular}[c]{@{}c@{}}Hidden\\ Trigger\end{tabular} } & \textbf{BP}      & \textbf{SIG}     & \textbf{ABE}     \\ \hline
\multirow{4}{*}{0.05}                                          & ANP              & 64.2\%                                                         & 51.6\%                 & 60.0\%          & 78.4\%          & 42.6\%          \\
                                                               & TopoTrigger      & 73.2\%                                                         & 74.0\%                 & 72.6\%          & \textbf{84.2\%} & 70.2\%          \\
                                                               & UNICORN          & 54.6\%                                                         & 75.2\%        & 74.0\%          & 86.8\%          & 73.0\%          \\
                                                               & CatchBackdoor    & \textbf{78.8\%}                                                & \textbf{77.2\%}                 & \textbf{72.0\%} & 83.6\%          & \textbf{76.2\%} \\ \hline
\multirow{4}{*}{0.01}                                          & ANP              & 66.2\%                                                         & 57.8\%                 & 63.4\%          & 75.0\%          & 47.8\%          \\
                                                               & TopoTrigger      & 72.4\%                                                         & 70.0\%                 & 71.8\%          & 85.8\% & 74.6\%          \\
                                                               & UNICORN          & \textbf{78.6\%}                                                         & \textbf{75.4\%}        & 71.2\%          & 84.0\%          & 75.8\%          \\
                                                               & CatchBackdoor    & 74.8\%                                                & 77.0\%                 & \textbf{72.6\%} & \textbf{86.2\%}          & \textbf{78.6\%}\\ \bottomrule
\end{tabular}
}
\label{Rate}
\end{table}

\subsection{Advanced attacks and large-scale dataset}
We add experiments on detecting WaNet~\cite{nguyen2021wanet}, LIRA~\cite{doan2021learnable}, and Marksman~\cite{doan2022marksman} on TinyImageNet under one-to-one attack setting. 
Table~\ref{tab:advanced_attacks} shows
CatchBackdoor effectively handles WaNet and LIRA attacks but struggles with the Marksman attack, 
likely due to Marksman's multi-attack target, which creates multiple trojan paths and complicates detection by CatchBackdoor. 
The larger number of classes does not significantly affect the performance of CatchBackdoor, showing its robustness.
\begin{table}[htbp]
\huge
\centering
\caption{Detection on advanced attacks on TinyImageNet. }
\resizebox{0.9\linewidth}{!}{
\begin{tabular}{cccccccc}
\toprule
\textbf{Models} & \textbf{Methods} & \textbf{BadNets} & \textbf{Hidden Trigger} & \textbf{ABE} & \textbf{WaNet} & \textbf{Lira} & \textbf{Marksman} \\ \hline
\multirow{4}{*}{VGG19} & ANP & 82.2\% & 57.6\% & N & N & N & N \\
 & TopoTrigger & 86.2\% & 70.4\% & 52.4\% & 61.0\% & N & 54.0\% \\
 & UNICORN & \textbf{90.0\%} & N & 58.6\% & 72.4\% & \textbf{76.2\%} & \textbf{56.2\%} \\
 & CatchBackdoor & 89.6\% & \textbf{73.8\%} & \textbf{68.0\%} & \textbf{78.6\%} & 75.2\% & N \\ \bottomrule
\end{tabular}
}
\label{tab:advanced_attacks}
\end{table}

\subsection{Random baseline analysis}
We select 50 benign examples per class and then set the neuron values of the original top-1 neural path to 0. We randomly select neurons to create a new top-1 neural path. Results of the prediction consistency rate are in Table~\ref{cr}. The random baseline only shows the consistency rate about 10\%, indicating the neural path controls model predictions.

\begin{table}[htbp]
\huge
\centering
\caption{Consistency rate after randomly replacing neural path}
\resizebox{0.6\linewidth}{!}{
\begin{tabular}{cccccccccccc}
\toprule
\multirow{2}{*}{Datasets} & \multirow{2}{*}{Models} & \multicolumn{10}{c}{Consistency Rate}                                                             \\ \cline{3-12} 
                          &                         & 0 & 1 & 2 & 3 & 4 & 5 & 6 & 7 & 8 & 9 \\ \hline
MNIST                     & LeNet-5                 & 8\%     & 6\%     & 12\%    & 6\%     & 8\%     & 16\%    & 10\%    & 4\%     & 4\%     & 10\%    \\
CIFAR-10                  & AlexNet                 & 2\%     & 14\%    & 6\%     & 2\%     & 12\%    & 8\%     & 0\%     & 6\%     & 18\%    & 6\%     \\ \bottomrule
\end{tabular}
}
\label{cr}
\end{table}

\section{Investigation of Neuron Behaviors}\label{neuron_behaviors} 
To further demonstrate our assumption, we compare the activation value and frequency between benign, trojaned and reversed examples for interpretation.

\begin{figure}[htbp]
\centering
    \subfloat[Activation value]{
        \includegraphics[width=0.6\linewidth]{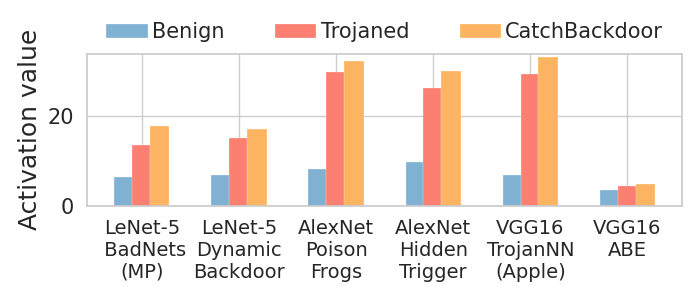}}\\
    \subfloat[Frequency]{
        \includegraphics[width=0.6\linewidth]{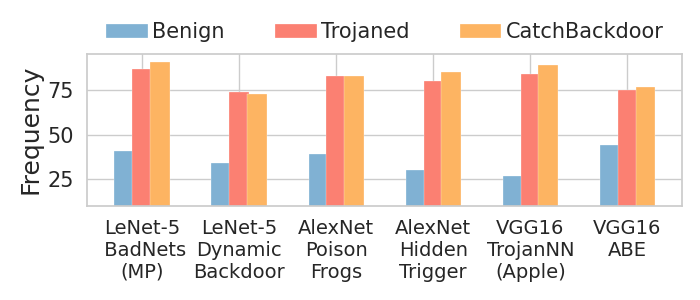}}
\caption{The activation value and frequency activated by benign examples, trojaned examples, and testing examples by CatchBackdoor.}
\label{mav}
\end{figure}

\begin{figure}[htbp]
\centering
\includegraphics[width=0.68\linewidth]{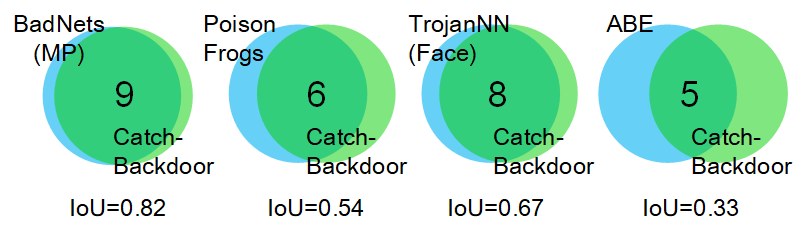}
\caption{The overlap of neurons activated by trojaned examples and CatchBackdoor's reversed examples in different trojaned models.}
\label{iou}

\end{figure}

\textbf{Implementation details.} (1) We randomly choose six models with trojan rate=0.1. 100 benign examples, trojaned examples and reversed examples by CatchBackdoor are input to the model. Then we calculate and compare the top-1 neuron contribution and frequency in the penultimate layer by different inputs. (2) We further compare the overlap of top-10 critical neurons activated by 100 trojaned examples and CatchBackdoor's reversed examples. AlexNet of CIFAR-10 is adopted and all models are trojaned with trojan rate=0.1. We calculate the Intersection over union (IoU) value of top-10 neurons in the penultimate layer. It is calculated as $IoU(X,X')=\frac{N(X)\cap N(X') }{N(X)\cup N(X')}$, where $X$ and $X'$ denote two batches of input examples. $N$ denotes the neurons activated by $X$ and $X'$, respectively. IoU of the activated neurons measures the consistency of neuron activation state. A greater value denotes higher consistency. (3) Fig.~\ref{mav} shows the results of activation value and frequency on different trojaned models. Fig.~\ref{iou} show the overlap of neurons. Each circle denotes 10 neurons. The corresponding IoU values between trojaned examples and CatchBackdoor's reversed examples are attached below.

\textbf{Results and analysis.}
From Fig.~\ref{mav}, we can see that the benign and trojaned examples show great differences in both activation value and activated frequency. 
Consistent with our assumption, neurons' activation values increase significantly when backdoors are triggered by trojaned examples. The activation value and frequency stimulated by CatchBackdoor generated examples are close to but higher than those by trojaned examples, which will trigger the trojaned behavior. Besides, the size and location of triggers have little effect on the activation value. Fig.~\ref{iou} further manifests that CatchBackdoor generates reversed examples that activate similar neurons as the trojaned examples. For instance, on four trojaned models, more than 50\% of top-10 neurons of CatchBackdoor are overlapped with trojaned examples, and these neurons are frequently chosen. By identifying similar neurons with trojaned examples during fuzzing, trojaned behaviors can finally be triggered. This explains the capability of finding potential backdoors by CatchBackdoor.

\section{Visualizations\label{Visualizations}}
In order to provide an intuitive understanding of CatchBackdoor, we provide visualization analysis in both input feature space and high-dimensional representative feature space in hidden layers. 

\begin{figure}[htbp]
\centering
\vspace{-0.3cm}
    \subfloat[BadNets (MP)]{
        \includegraphics[width=0.14\linewidth]{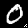}
        \includegraphics[width=0.14\linewidth]{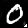}}
    \subfloat[BadNets (IP)]{
        \includegraphics[width=0.14\linewidth]{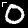}
        \includegraphics[width=0.14\linewidth]{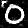}
        }
    \subfloat[Poison frogs]{   
    \includegraphics[width=0.14\linewidth]{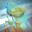}
    \includegraphics[width=0.14\linewidth]{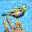}}\\
     \subfloat[TrojanNN (Face)]{
     \includegraphics[width=0.14\linewidth]{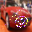}
     \includegraphics[width=0.14\linewidth]{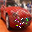}
         }
    \subfloat[TrojanNN (Apple)]{
    \includegraphics[width=0.14\linewidth]{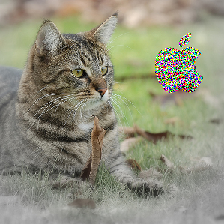}
    \includegraphics[width=0.14\linewidth]{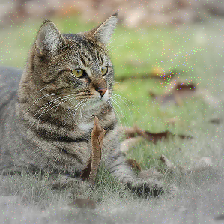}
    }
    \subfloat[ABE]{
        \includegraphics[width=0.14\linewidth]{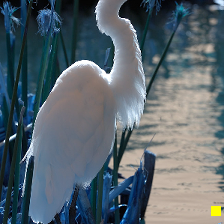}
        \includegraphics[width=0.14\linewidth]{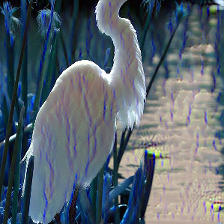}}\\
\caption{Visualizations of trojaned examples (left) and reversed examples (right) that change model prediction labels.}
\label{Triggers_Testing_Reverse}

\end{figure}

Fig.~\ref{Triggers_Testing_Reverse} shows trojaned examples and our generated reversed examples under different trojan attacks, i.e., BadNets, Poison frogs, TrojanNN and ABE. LeNet-5 of MNIST, AlexNet of CIFAR-10 and VGG16 of a-ImageNet are used for visualization. 
A high visual similarity in appearance and trigger position can be observed from the figures. This well verifies the effectiveness of CatchBackdoor in reversing triggers. Based on the critical trojan path, CatchBackdoor can target more label-related pixels frequently used for classification.

\makeatletter

\begin{figure}[htbp]
\centering
    \includegraphics[width=0.6\linewidth]{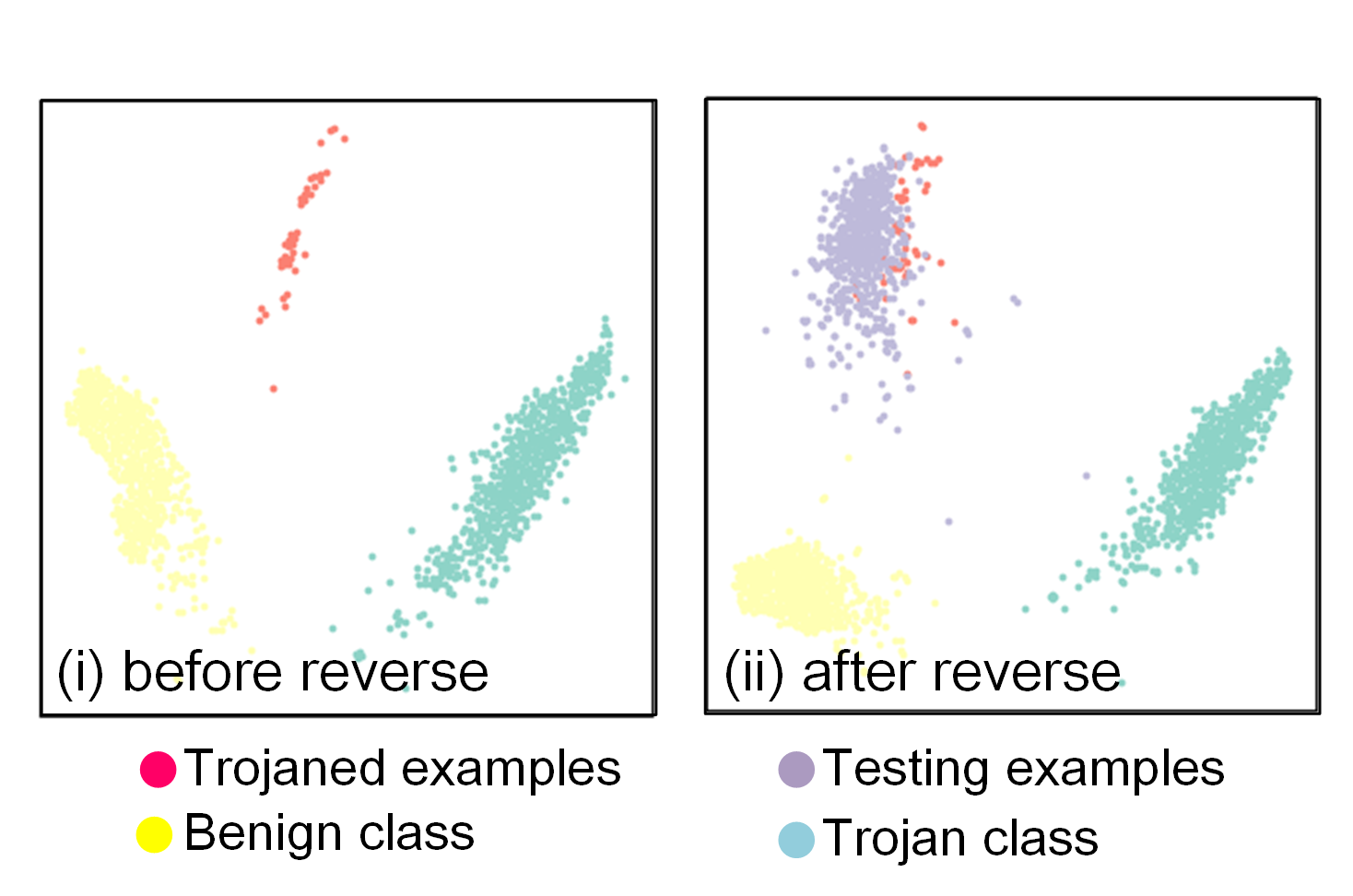}
\caption{The t-SNE visualization of high-dimensional features of benign, trojan, and reversed examples by CatchBackdoor. }

\label{tsne testing}
\end{figure}

Furthermore, we visualize high-dimensional features via t-SNE on LeNet-5 model trojaned by BadNets (MP) with trojan rate 0.1. The results are shown in Fig.~\ref{tsne testing}, where the feature distribution of trojaned examples, reversed examples, benign and trojaned class examples are represented by red, purple, yellow, and green clusters, respectively. It can be seen that the well-trained model can distinguish different labels as clusters separate from each other. Reversed examples by CatchBackdoor can serve as trojaned examples to trigger backdoors, as red cluster overlap the purple one. As a result, reversed examples can trigger misclassifications due to backdoors. 

\section{Detection for Online Model Zoo}\label{model_zoo_res} 
\begin{figure}[htbp]
\centering
    \subfloat[age]{
        \includegraphics[width=0.45\linewidth]{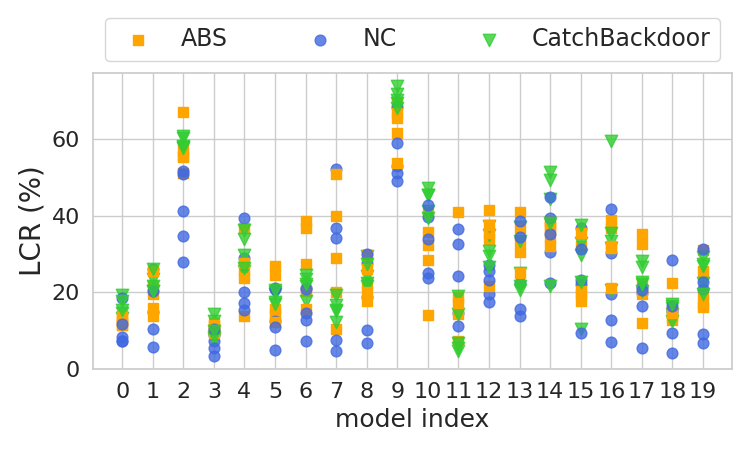}
    }
    \hfill
    \subfloat[gender]{
        \includegraphics[width=0.45\linewidth]{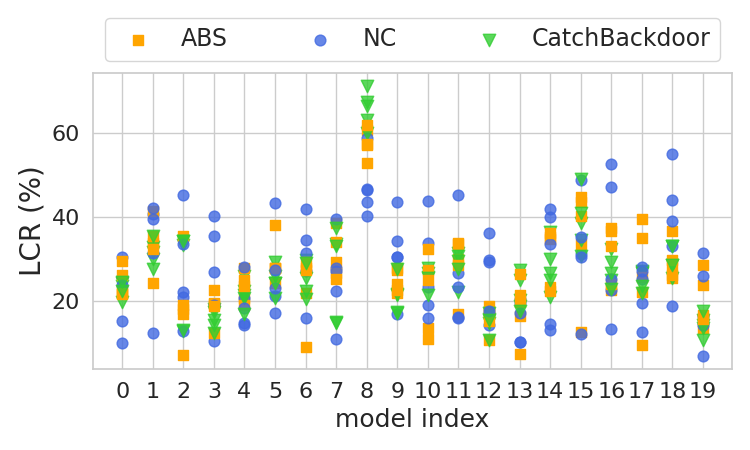}
    }
\caption{LCR for Model Zoo Models of age and gender classification tasks. Twenty models with the highest LCR are shown.}
\label{modelzoo}
\end{figure}

We apply CatchBackdoor to test pre-trained models on Caffe Model Zoo. 100 pre-trained models are downloaded for age classification and 100 for gender classification. We run detection for 5 rounds to evaluate the potential risks of the models. 1,000 randomly selected examples from each label are reorganized for test sets. We calculate LCR on CatchBackdoor along with the two baselines, ABS and NC. Fig.~\ref{modelzoo} shows the results of 20 models with the highest LCR values of each classification task, where scatters of orange, blue and green denote the LCR of ABS, NC and CatchBackdoor, respectively. Models with larger LCR are more likely to be trojaned (naturally\footnote{A natural backdoor is not implanted manually. Instead, the model may have a specific pattern that can trigger a similar behavior to that of a backdoor trigger.} or manually).

According to the figure, compared with baselines, reversed examples generated by CatchBackdoor show higher LCR than baselines in most cases. Three methods show the same trend during the testing that the No.2 and 9 model in age classification and No.8 in gender are likely to be trojaned, with higher LCR. In the five rounds of testing, all models with potential deflects show LCR higher than 50\% by CatchBackdoor so that they can be easily detected. This indicates CatchBackdoor can more accurately discover the potential backdoors inside the model. Most models show low LCR among the three methods, so they are prone to be safe for downstream consumers.

\section{Detection against Adaptive Attacks}\label{Adaptive_Attacks}
We design two types of adaptive trojan attacks, i.e., standardized adaptive attack (S-AA) and imitation adaptive attack (I-AA) and further evaluate CatchBackdoor on them. 

Suppose $X$ is the benign training set with its corresponding label pair set $Y$. $\bar{X}$ and $\bar{Y}$ denote the trojan set and trojan label set for training. $X^*= X\bigcup \bar{X} $, $Y^*= Y\bigcup \bar{Y}$. Here we give formal definitions of S-AA and I-AA.

\noindent \textbf{\textcircled{1} S-AA.} S-AA aims to minimize the standard deviation of neuron contributions in the same layer during backdoor training while maintaining high benign accuracy. For each $x^*\in X^*, y^*\in Y^*$, the loss function of training the trojaned model is defined as:
\begin{equation}
    \mathop{\arg\min}\limits_{\theta^{*}}~ \mathcal{L}(x^*,y^*,\theta^{*}) + \delta^2(\bigcup_{j=1}^{k}\xi_{n_{l,j}}(x^*)) 
\end{equation}

where $\mathcal{L}$ is the cross-entropy loss of the trojaned model with parameters $\theta^{*}$. $\xi_{n_{l,j}}(\cdot)$ represents neuron contribution of the $j$-th neuron in the $l$-th layer by input $x^*$.  $\delta^2(\cdot)$ calculates the standard deviation of neuron activations in the $l$-th layer. $k$ denotes the number of chosen neurons. Here, the penultimate layer is chosen and $k$ is set to 5.

\noindent \textbf{\textcircled{2} I-AA.} Unlike S-AA, I-AA achieves the trojan attack by controlling trojan neuron activation behaviors similar to benign ones. For each trojaned example $\bar{x} \in \bar{X}$, benign input $x \in X$, the loss function is defined as:
\begin{equation}
    \mathop{\arg\min}\limits_{\theta^{*}}~ \gamma  \mathcal{L}(x^*, y^*,\theta^{*}) + (1\!\!-\!\!\gamma) M\!S\!E(\bigcup_{i=1}^{l-1}\bigcup_{j=1}^{k} \xi_{\bar{n}_{i,j}}(\bar{x}), \bigcup_{i=1}^{l-1}\bigcup_{j=1}^{k} \xi_{n_{i,j}}(x))
\end{equation}

The training configurations are shown in~\ref{Adaptive_conf}. For all adaptive attacks, We generate 1,000 testing examples for calculating LCR.  We set both $k_1$ and $k_2$ are set to 5.

\begin{table}[htbp]
\centering
\large
\caption{The training configurations of adaptive attacks.}
\resizebox{0.72\linewidth}{!}{
\begin{tabular}{ccccccc}
\toprule
\multirow{2}{*}{\textbf{Datasets}} & \multirow{2}{*}{\textbf{Models}} & \multirow{2}{*}{\textbf{acc}} & \multicolumn{4}{c}{\textbf{Configurations}}                                                  \\ \cline{4-7} 
                                   &                                  &                               & \textbf{batch\_size}    & \textbf{\#Epochs} & \textbf{patch\_size}    & \textbf{trojan rate} \\ \hline
\multirow{2}{*}{MNIST}             & \multicolumn{1}{c}{LeNet-4}     & 97.20\%                        & \multicolumn{1}{c}{32} & 10                & 1$\times$1 & 0.1                 \\
                                   & LeNet-5                          & 97.50\%                        & 32                      & 10                & 1$\times$1 & 0.1                 \\ \hline
\multirow{2}{*}{CIFAR-10}          & AlexNet                          & 93.60\%                        & 32                      & 20                & 3$\times$3 & 0.1                \\
                                   & ResNet20                         & 93.00\%                        & 32                      & 20                & 3$\times$3 & 0.1                 \\ \hline
\multirow{2}{*}{a-ImageNet}     & VGG16                            & 92.10\%                        & 32                      & 40                & 7$\times$7 & 0.1                 \\
                                   & \multicolumn{1}{c}{VGG19}       & 91.70\%                        & 32                      & 40                & 7$\times$7 & 0.1                 \\ \bottomrule
\end{tabular}}
\label{Adaptive_conf}
\end{table}

\section {Model Hardening by Retraining DNNs\label{robust improve}}
We use reversed examples for retraining and calculate the robustness improvement.

\begin{figure}[htbp]
\centering
    \includegraphics[width=0.7\linewidth]{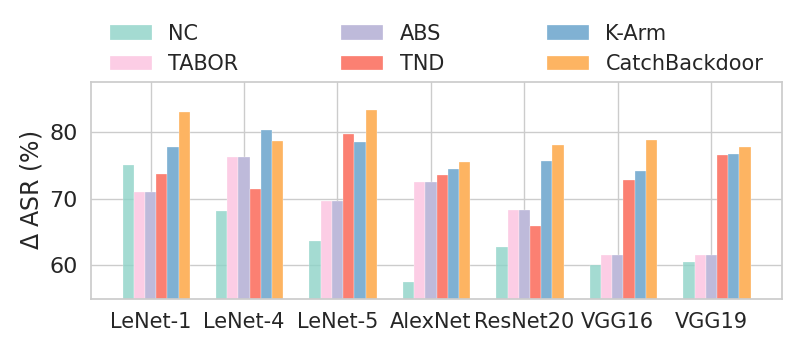}
\caption{The $\Delta$ ASR of BadNets before and after fine-tuning on different models.}
\label{retrain-trojan}
\end{figure}

\textbf{Implementation details.} (1) The datasets and models we used are the same as those mentioned before. (2) BadNets is used to construct trojaned models, with trojan rate=0.1. (3) 2,000 reversed examples from the same benign seeds are generated by five baselines and CatchBackdoor. Then they are 1: 1 mixed with reversed examples to form training data. The original models are fine-tuned for 20 epochs with $batch\_size=32$ while maintaining similar classification accuracy. (4) The $\Delta$ ASR before and after the retraining measured by 500 trojaned examples are shown in Fig.~\ref{retrain-trojan}.

\textbf{Results and analysis.} From the figure, we can conclude that retraining can effectively reduce the ASR of trojan attacks. Compared with baselines, models fine-tuned by CatchBackdoor's reversed examples are more robust against trojan attacks. CatchBackdoor can cover more diverse corner cases that may lead to misclassification. By retaining those examples, the robustness of models is better enhanced.

\section {Time Complexity \label{time}}
In this section, we analyze the efficiency of CatchBackdoor, i.e., reverse-engineering time cost.

\textbf{Implementation details.} (1) We measure average running time for generating 100 reversed examples from the same benign seeds by CatchBackdoor and baselines. We run each method 3 times, and the minimal one is identified as the final result. (2) For CatchBackdoor, $k_1,k_2=3$. (3) The running time (seconds) is shown in Table~\ref{Testing Time Comparison}.

\begin{table}[htbp]
\huge
\centering
\caption{Time (seconds) taken to reverse 100 examples.}
\resizebox{0.7\linewidth}{!}{
\begin{tabular}{cccccccc}
\toprule
\multirow{2}{*}{\textbf{Datasets}}       & \multirow{2}{*}{\textbf{Models}} & \multicolumn{6}{c}{\textbf{Methods}}                      \\
                               &                        & NC    & TABOR & ABS & TND & K-Arm & CatchBackdoor \\ \hline
\multirow{3}{*}{MNIST}         & LeNet-1                & 301.32   & 299.24   & 163.35 & 178.0 & 177.67   & 172.01           \\
                               & LeNet-4                & 517.34   & 520.47   & 174.50 & 192.82 & 186.37   & 178.55           \\
                               & LeNet-5                & 634.02   & 547.45   & 185.40 & 207.37 & 268.85   & 205.20           \\ \hline
\multirow{2}{*}{CIFAR-10}      & AlexNet                & 1378.37 & 1216.98 & 699.20 & 714.49 & 744.95   & 715.87           \\
                               & ResNet20              & 1715.20 & 1374.03 & 792.50 & 813.91 & 828.11   & 794.38           \\ \hline
\multirow{2}{*}{a-ImageNet} & VGG16                  & 1878.20 & 1465.59 & 819.95 & 827.30 & 903.05   & 874.38           \\
                               & VGG19                  & 1906.65 & 1483.33 & 863.67 & 894.22 & 959.20   & 939.84          \\ \bottomrule
\end{tabular}
}
\label{Testing Time Comparison}

\end{table}

\textbf{Results and analysis.} 
Firstly, we theoretically analyze the complexity of CatchBackdoor according to different steps.

In the step of benign path construction, we use a balanced tree data structure, which allows for quicker search, insert, and delete operations. So the computation complexity can be calculated as:

\begin{equation}
    T_{construct}\sim \mathcal{O}(N \times \log_{}{a})
\end{equation}
where $N$ denotes the total number of benign seeds, and $a$ is the total number of neurons in the model.

In the fuzzing step, we use a divide-and-conquer approach, MergeSort and run $S$ rounds of iterations per example to maximize the neuron contribution of each neuron in the benign path. So the computation complexity can be calculated as:
\begin{equation}
    T_{fuzzing}\sim \mathcal{O}(N \times \log_{}{b})
\end{equation}
where $b \textless a$ denotes the total number of critical neurons.

In the step of trigger reverse engineering, we implement binary search for identifying triggers. Then the computation complexity is as follows:
\begin{equation}
    T_{generation}\sim \mathcal{O}(N)
\end{equation}
Therefore, the total time complexity is: $\mathcal{O}(N)$.

Secondly, we analyze the efficiency of CatchBackdoor from the real running time. According to Table~\ref{Testing Time Comparison}, the running time of CatchBackdoor is acceptable. With the increasing model complexity, the time cost of CatchBackdoor increases due to the increase of the total number of neurons. In small datasets, CatchBackdoor is slightly inferior to ABS, but is still much faster than NC. The reason is that we search more layers and neurons, compared with ABS. Besides, CatchBackdoor fuzzs diverse inputs and monitors label changes while NC, TABOR, and TND must scan all the output labels one by one and construct the minimum perturbation for each label.

\section{Random noise triggers benign path\label{appendix_noise}}
\begin{figure*}[htbp]
\centering
\includegraphics[width=0.8\linewidth]{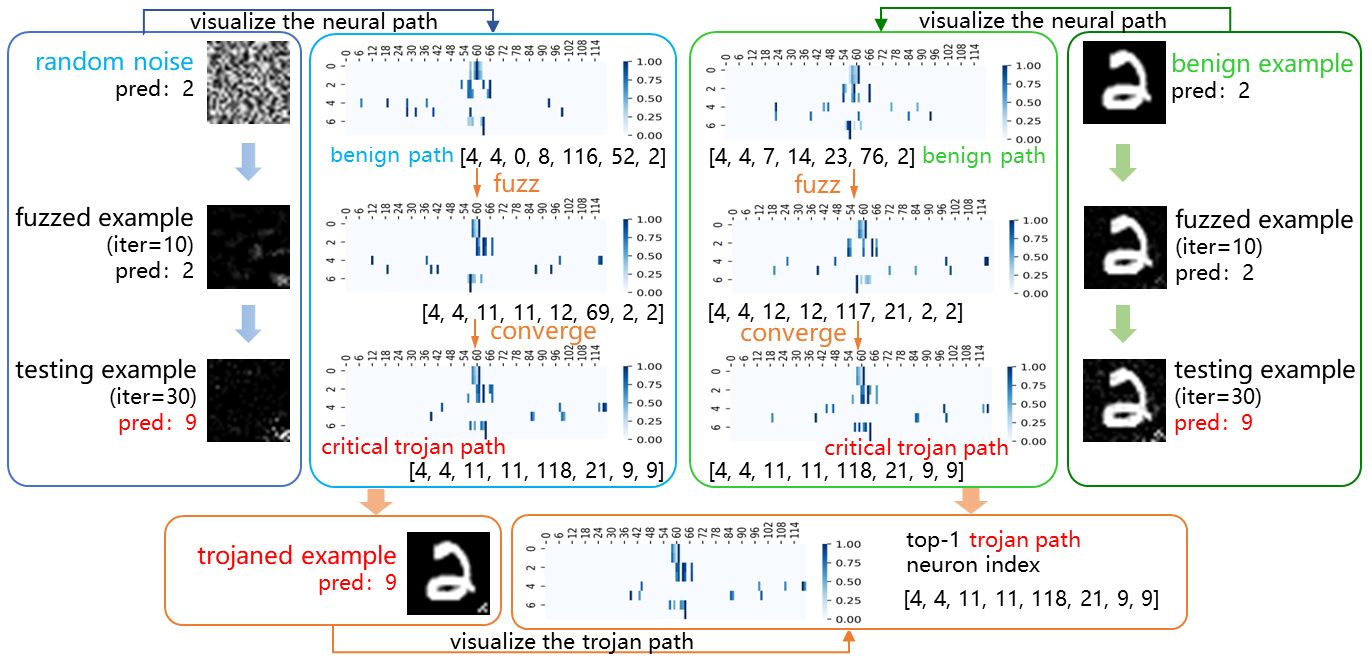}
\caption{Illustration of benign path, critical trojan path and trojan path activated by a random noise, a benign example and a trojaned example. Top-5 neural path of each example is shown via heatmap. The neuron index in top-1 neural path in each layer is presented. Dataset: MNIST; Model: LeNet-5.}
\label{noisecase}
\end{figure*}
We have conducted an experiment to prove the feasibility of using random noise to construct benign path when benign examples are unavailable, as shown in Figure~\ref{noisecase}. For random noise, the reversed example is generated by putting the reversed trigger on the black image for better visualization. LeNet-5 of MNIST trojaned by BadNet, is adopted for visualization. We set trojaned label=``9'', trojan rate=0.1, $batch\_ size=32$, $epoch=10$. ``iter'' refers to the current iteration number. Here we set the maximum iteration number $S$ to 30. In the heatmap, the contribution of neurons in the top-5 path after normalization to [0, 1] is presented. The color goes deeper as the value grows. The x-axis is the neuron index while the y-axis is the layer index. Gaussian distribution with $\mu=0$ and $\delta=1$ is selected for generating initial random noise.

To begin with, a random noise example fed into the trojaned model is classified as ``2'', with the top-1 benign path as [4, 4, 0, 116, 52, 2]. This is very different from the top-1 benign path activated by benign example with the same label ``2'', whose top-1 benign path is [4, 4, 7, 14, 32, 76, 2]. With the growing number of iterations, the trigger will be gradually reversed. When fuzzing finishes in 30th iteration, the fuzzed path finally converges to the critical neural path. Obviously, when starting from random noise or the benign example to construct benign paths, the same top-1 critical trojan path can be gained, which is also the same as that activated by the real trojaned example. This well verifies that random noise can be an option for starting fuzzing when benign training data is unavailable.

\section{Visualization of top-5 neural path\label{Visualizationtop-5}}
\begin{figure*}[htbp]

\makeatother
\centering
\captionsetup[subfloat]{labelsep=none,format=plain,labelformat=empty}
    \subfloat[(a1) BadNets-noise]{
        \includegraphics[width=0.4\linewidth]{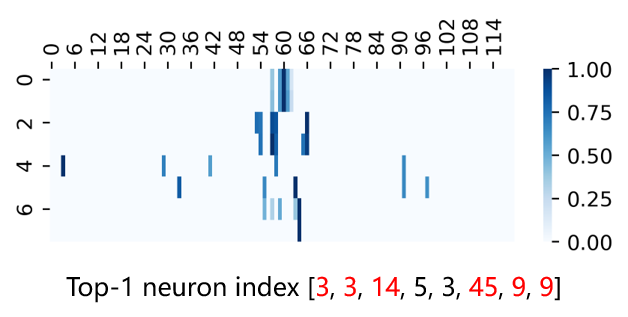}}
    \hspace{10pt}
    \subfloat[(a2) BadNets-trojaned example]{
        \includegraphics[width=0.4\linewidth]{figs/mnist_bd_noise.png}}\\
    \subfloat[(b1) Poison frogs-noise]{
        \includegraphics[width=0.4\linewidth]{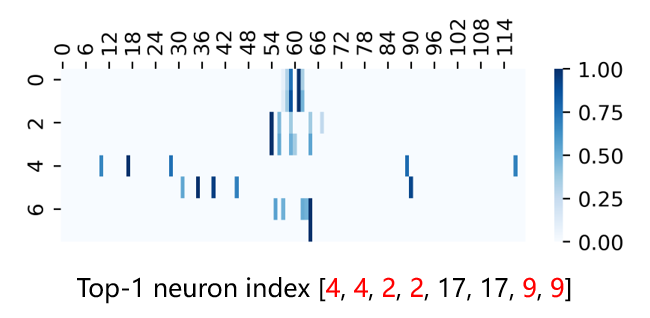}}
    \hspace{10pt}
    \subfloat[(b2) Poison frogs-trojaned example]{
        \includegraphics[width=0.4\linewidth]{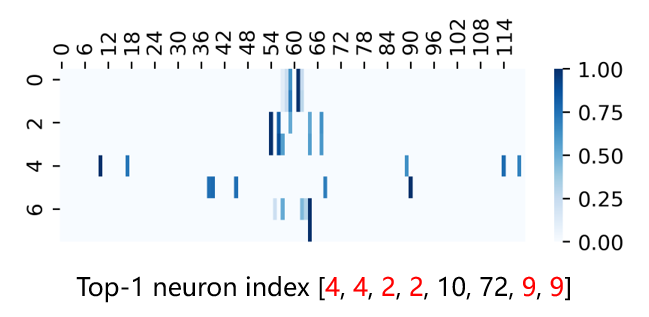}}\\
        \subfloat[(c1) TrojanNN (Face)-noise]{
        \includegraphics[width=0.4\linewidth]{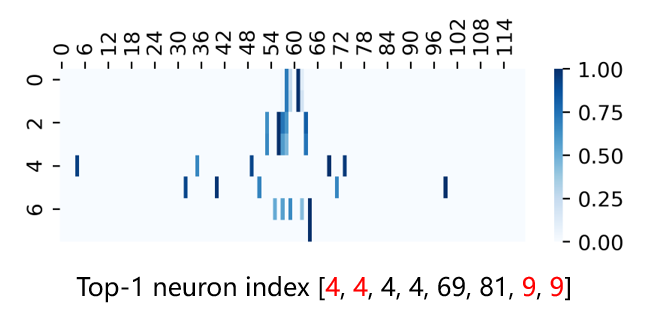}}
    \hspace{10pt}
    \subfloat[(c2) TrojanNN (Face)-trojaned example]{
        \includegraphics[width=0.4\linewidth]{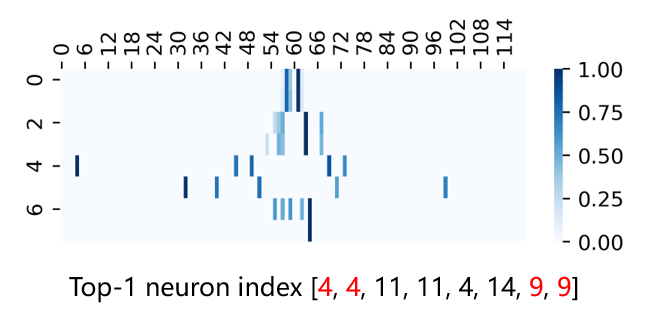}}\\
    \subfloat[(d1) ABE-noise]{
        \includegraphics[width=0.4\linewidth]{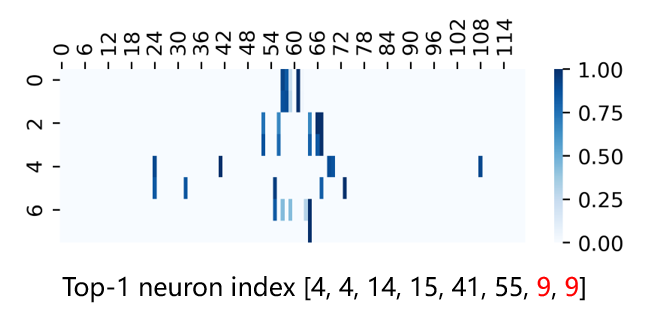}}
    \hspace{10pt}
    \subfloat[(d2) ABE-trojaned example]{
        \includegraphics[width=0.4\linewidth]{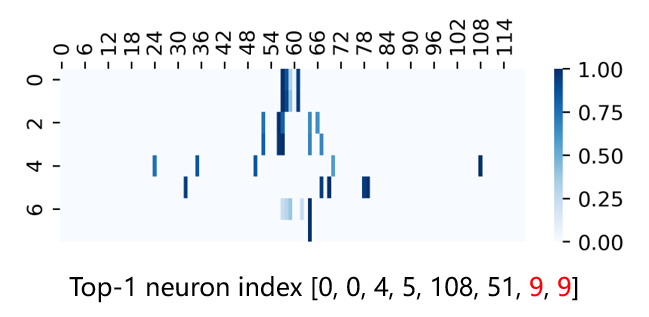}}\\
\caption{The heatmap of top-5 neural path of reversed examples fuzzed from random noise after 30 iterations and that of trojaned examples. The neuron index in top-1 neural path in each layer is presented. The same neuron index between two top-1 neural paths is highlight in red. The trojaned label of all attacks is ``9'', and the trojan rate is set to 0.1. Dataset: MNIST; Model: LeNet-5.} 
\label{mnist_noise}
\end{figure*}

We visualize the top-5 critical trojan path fuzzed from random noise, and trojan path activated by trojaned examples after 30 iterations via heatmap. We adopt 4 different attacks including BadNets, Poison frogs, TrojanNN, and ABE at 3 models. The trojaned label of all attacks is ``9'' with trojan rate=0.1. The visualization results of LeNet-5 of MNIST, AlexNet of CIFAR-10, VGG19 of a-ImageNet are shown in Fig.~\ref{mnist_noise}, Fig.~\ref{cifar_noise2} and Fig.~\ref{image_noise2}, respectively.

\makeatletter

\makeatletter
\begin{figure*}[htbp]
\makeatother
\centering
\captionsetup[subfloat]{labelsep=none,format=plain,labelformat=empty}
    \subfloat[(a1) BadNets-noise]{
        \includegraphics[width=1\linewidth]{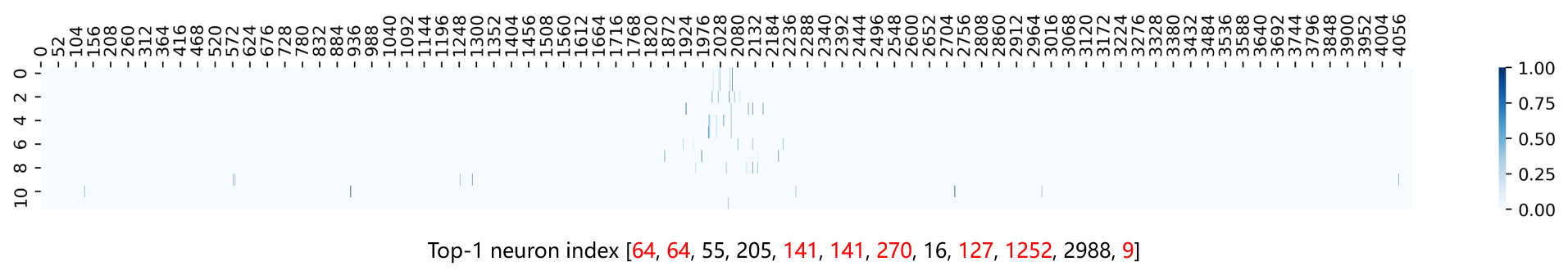}}\\
    \subfloat[(a2) BadNets-trojaned example]{
        \includegraphics[width=1\linewidth]{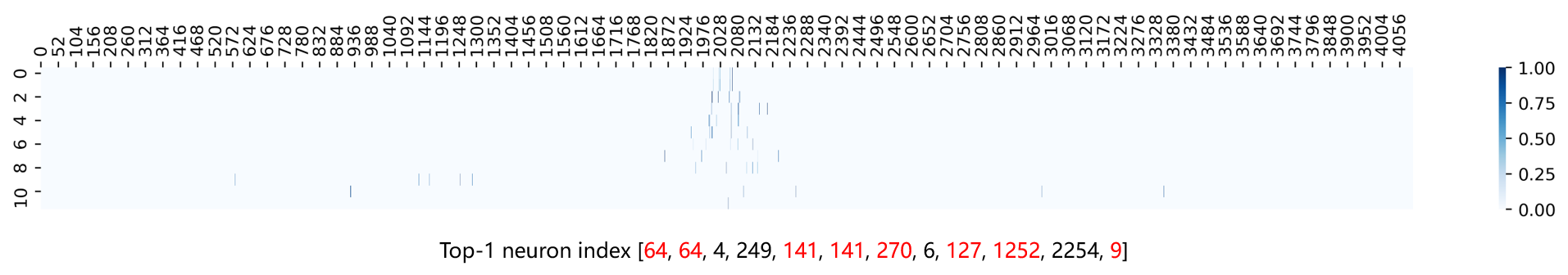}}\\
    \subfloat[(b1) Poison frogs-noise]{
        \includegraphics[width=1\linewidth]{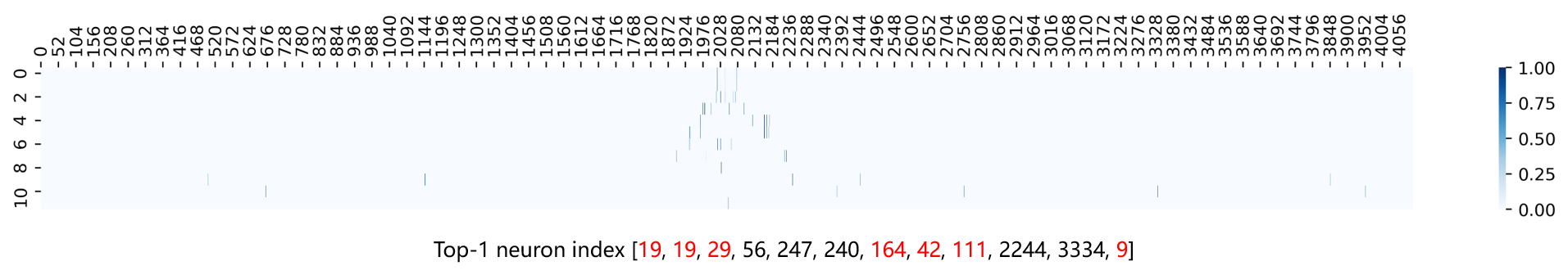}}\\
    \subfloat[(b2) Poison frogs-trojaned example]{
        \includegraphics[width=1\linewidth]{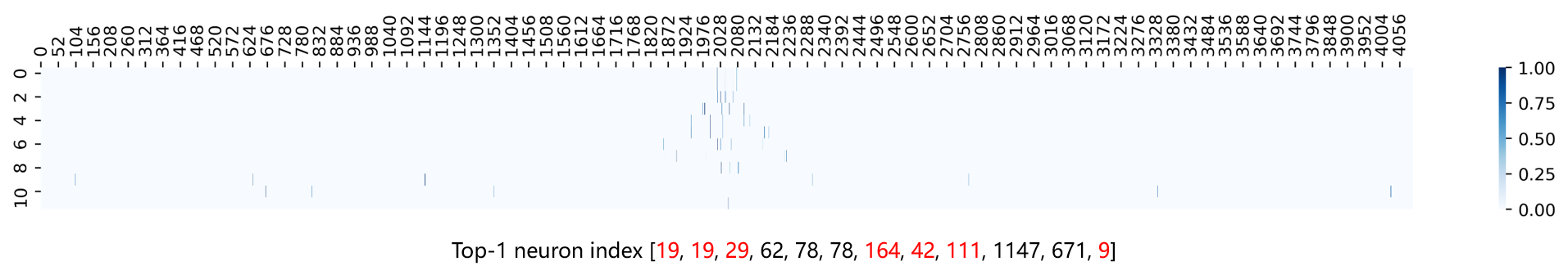}}\\
    \subfloat[(c1) TrojanNN (Face)-noise]{
        \includegraphics[width=1\linewidth]{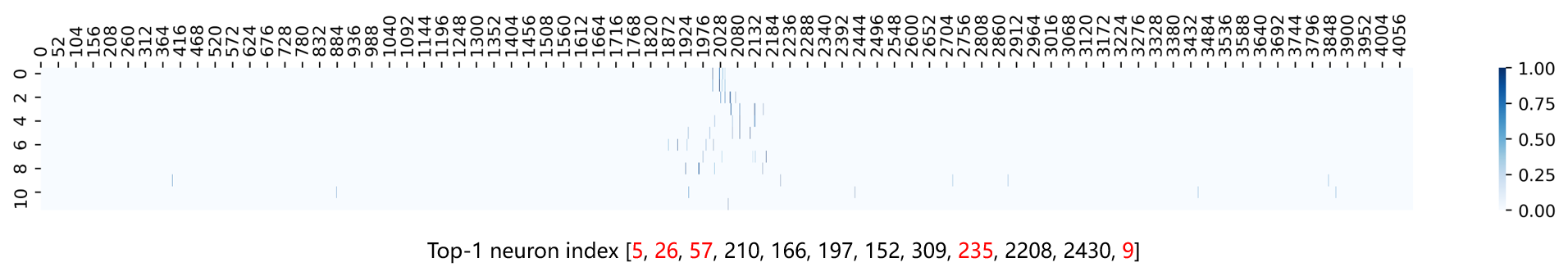}}\\
    \subfloat[(c2) TrojanNN (Face)-trojaned example]{
        \includegraphics[width=1\linewidth]{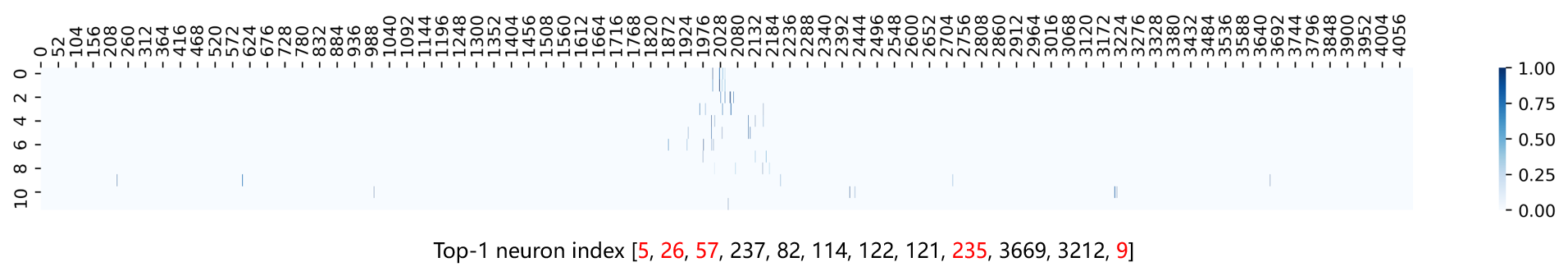}}\\
            \subfloat[(d1) ABE-noise]{
        \includegraphics[width=1\linewidth]{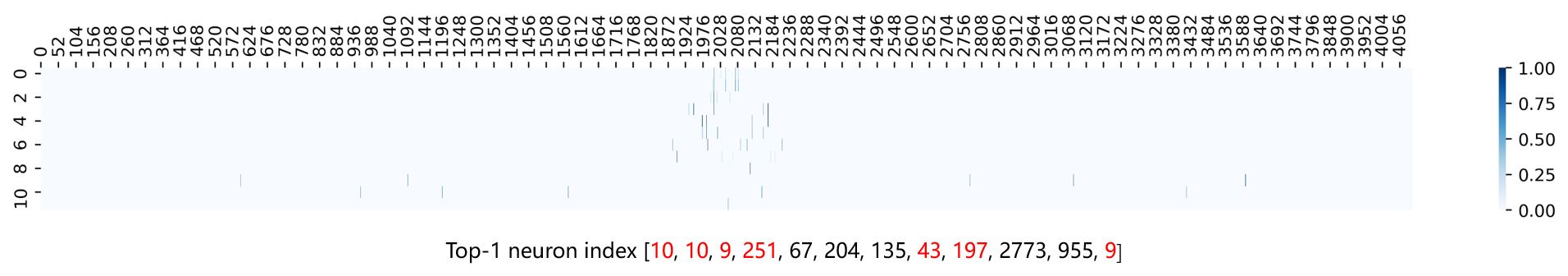}}\\
\label{cifar_noise1}
\end{figure*}
\clearpage
\begin{figure*}[htbp]
\makeatother
\addtocounter{figure}{-1} 
\centering
\captionsetup[subfloat]{labelsep=none,format=plain,labelformat=empty}   
    \subfloat[(d2) ABE-trojaned example]{
        \includegraphics[width=1\linewidth]{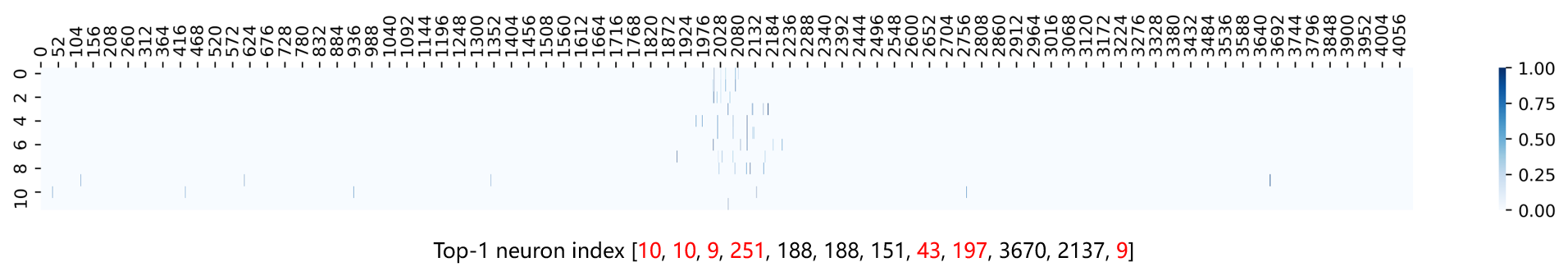}}\\
\caption{The heatmap of the top-5 neural path of the reversed examples fuzzed from random noise after 30 iterations and that of the trojaned examples. The neuron index in top-1 neural path in each layer is presented. The same neuron index between two top-1 neural paths is highlighted in red. The trojaned label of all attacks is ``9'', and the trojan rate is set to 0.1. Dataset: CIFAR-10; Model: AlexNet.}
\label{cifar_noise2}
\end{figure*}
\newpage

\makeatletter
\begin{figure*}[htbp]
\makeatother
\centering
\captionsetup[subfloat]{labelsep=none,format=plain,labelformat=empty}    
    \subfloat[(a1) BadNets-noise]{
        \includegraphics[width=1\linewidth]{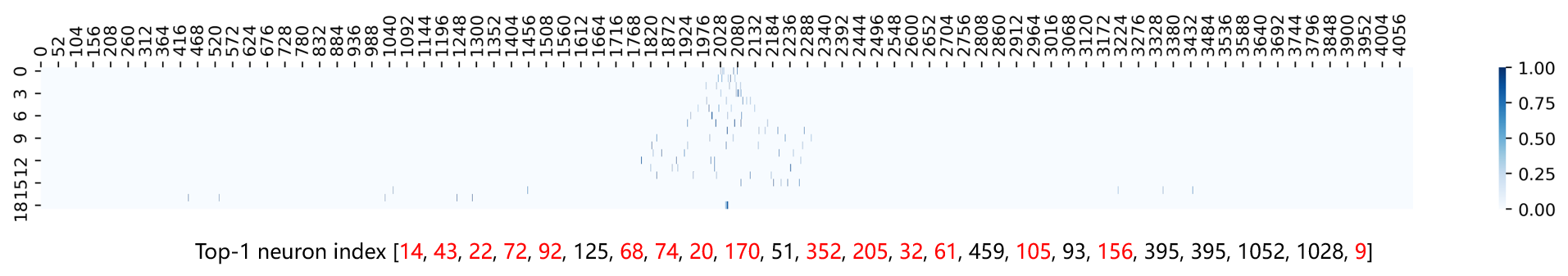}}\\
    \subfloat[(a2) BadNets-trojaned example]{
        \includegraphics[width=1\linewidth]{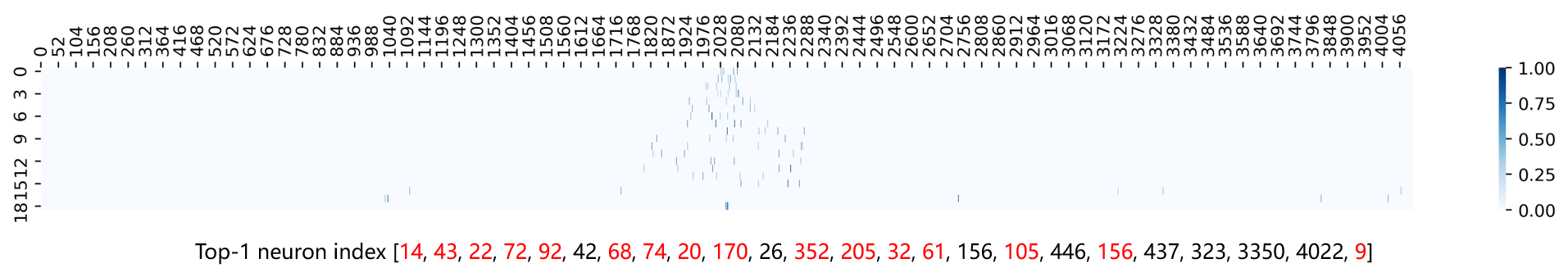}}\\
    \subfloat[(b1) Poison frogs-noise]{
        \includegraphics[width=1\linewidth]{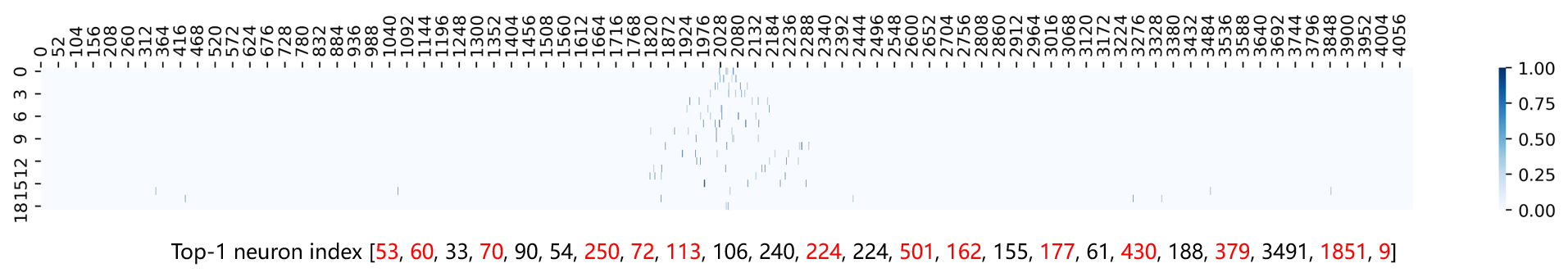}}\\
    \subfloat[(b2) Poison frogs-trojaned example]{
        \includegraphics[width=1\linewidth]{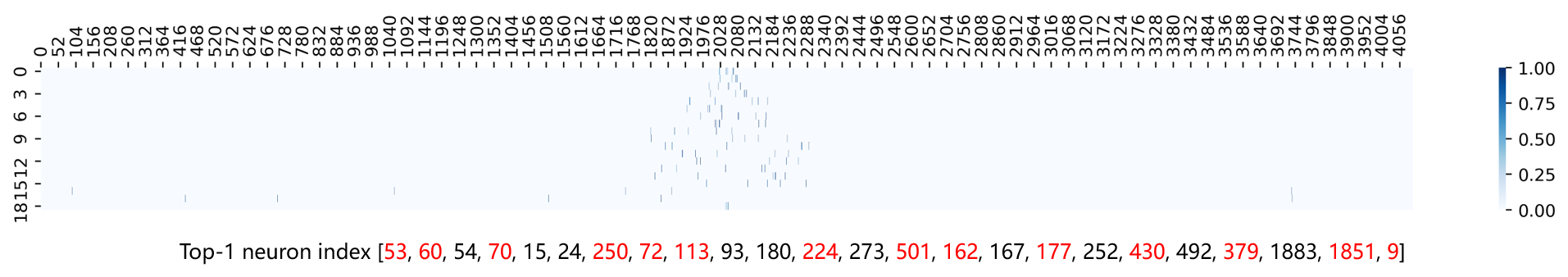}}\\
    \subfloat[(c1) TrojanNN (Face)-noise]{
        \includegraphics[width=1\linewidth]{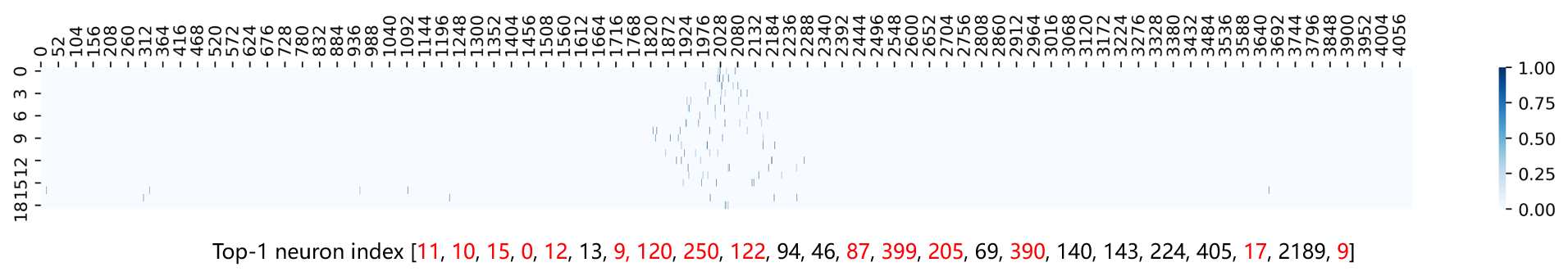}}\\
    \subfloat[(c2) TrojanNN (Face)-trojaned example]{
        \includegraphics[width=1\linewidth]{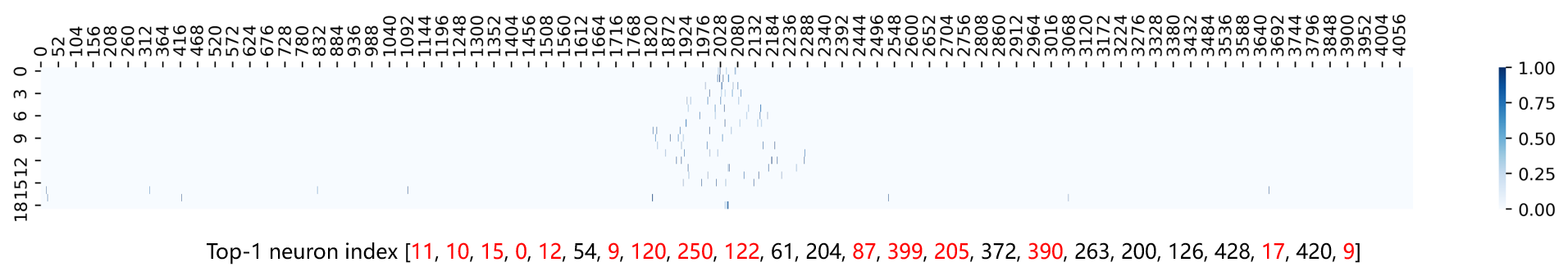}}\\
        \subfloat[(d1) ABE-noise]{
        \includegraphics[width=1\linewidth]{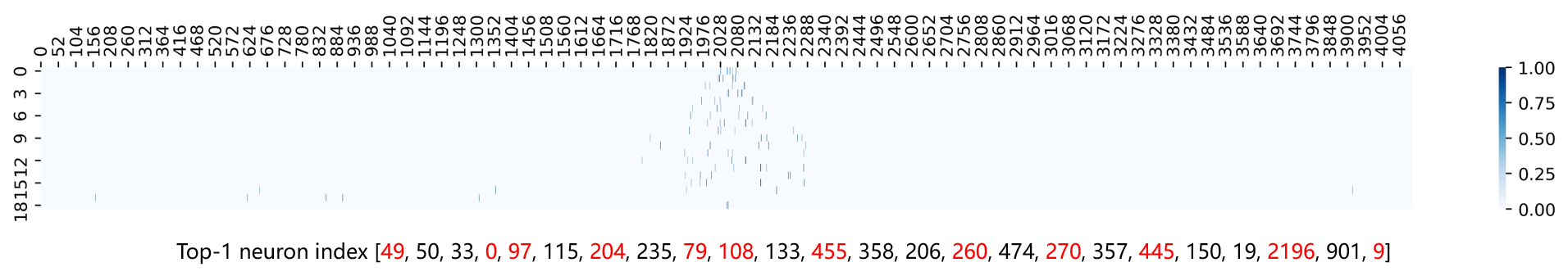}}\\
\label{image_noise1}
\end{figure*}
\clearpage
\addtocounter{figure}{-2} 
\makeatletter
\begin{figure*}[htbp]
\makeatother
\addtocounter{figure}{1} 
\captionsetup[subfloat]{labelsep=none,format=plain,labelformat=empty}
    \subfloat[(d2) ABE-trojaned example]{
        \includegraphics[width=1\linewidth]{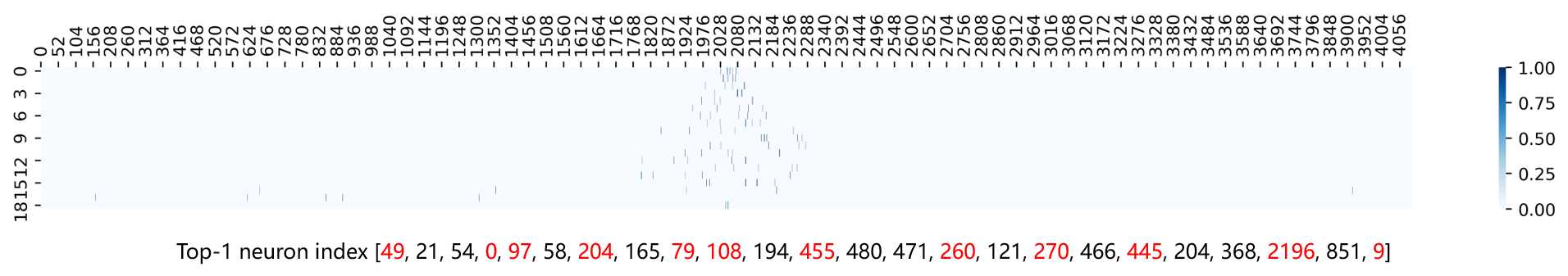}}\\
\caption{The heatmap of the top-5 neural path of the reversed examples fuzzed from random noise after 30 iterations and that of the trojaned examples. The neuron index in top-1 neural path in each layer is presented. The same neuron index between two top-1 neural paths is highlighted in red. The trojaned label of all attacks is ``9'', and the trojan rate is set to 0.1. Dataset: a-ImageNet; Model: VGG19.}
\label{image_noise2}
\end{figure*}



\end{document}